\documentclass[twocolumn]{aastex631}

\newcommand{\aref}[1]{\hyperref[#1]{Appendix~\ref{#1}}}

\newcommand{\kepler}{\textit{Kepler}}
\newcommand{\tess}{\textit{TESS}}
\newcommand{\nAll}{232,705}  
\newcommand{\nSamp}{199,412}  
\newcommand{\nSingle}{152,993}  
\newcommand{\nMult}{79,712}  
\newcommand{\nMultSamp}{69,253}  
\newcommand{\nSine}{68,497}  
\newcommand{\nSineSig}{30,370}  
\newcommand{\nSineLow}{38,127}  
\newcommand{\nDsine}{10,887}  
\newcommand{\nACF}{4,662}  
\newcommand{\nSineMult}{19,118}  
\newcommand{\nDsineMult}{3,907}  
\newcommand{\nACFmult}{1,900}  

\newcommand{\nTotal}{84,046} 
\newcommand{\nTotalSig}{45,919} 

\newcommand{\Lsun}{\mbox{L$_{\odot}$}}
\newcommand{\Rsun}{\mbox{R$_{\odot}$}}

\usepackage{lipsum}

\shorttitle{TESS Stellar Variability Catalog}
\shortauthors{Fetherolf et al.}

\begin{document}

\title{Variability Catalog of Stars Observed During the TESS Prime Mission}

\correspondingauthor{Tara Fetherolf}
\email{tara.fetherolf@gmail.com}

\author[0000-0002-3551-279X]{Tara Fetherolf}
\altaffiliation{UC Chancellor's Fellow}
\affiliation{Department of Earth and Planetary Sciences, University of California Riverside, 900 University Avenue, Riverside, CA 92521, USA}

\author[0000-0002-3827-8417]{Joshua Pepper}
\affiliation{Department of Physics, Lehigh University, 16 Memorial Drive East, Bethlehem, PA 18015, USA}

\author[0000-0003-0447-9867]{Emilie Simpson}
\affiliation{Department of Earth and Planetary Sciences, University of California Riverside, 900 University Avenue, Riverside, CA 92521, USA}
\affiliation{SETI Institute, Mountain View, CA  94043, USA}

\author[0000-0002-7084-0529]{Stephen R. Kane}
\affiliation{Department of Earth and Planetary Sciences, University of California Riverside, 900 University Avenue, Riverside, CA 92521, USA}

\author[0000-0003-4603-556X]{Teo Mo\v{c}nik}
\affiliation{Gemini Observatory/NSF's NOIRLab, 670 N. A'ohoku Place, Hilo, HI 96720, USA}

\author[0009-0004-1179-7438]{John Edward English}
\affiliation{Department of Computer Science, University of California Irvine, 6210 Donald Bren Hall, Irvine, CA 92697, USA}

\author[0000-0002-0865-3650]{Victoria Antoci}
\affiliation{DTU Space, National Space Institute, Technical University of Denmark, Elektrovej 328, DK-2800 Kgs. Lyngby, Denmark}
\affiliation{Stellar Astrophysics Centre, Department of Physics and Astronomy, Aarhus University, Ny Munkegade 120, DK-8000 Aarhus C, Denmark}

\author[0000-0001-8832-4488]{Daniel Huber}
\affiliation{Institute for Astronomy, University of Hawai`i, 2680 Woodlawn Drive, Honolulu, HI 96822, USA}

\author[0000-0002-4715-9460]{Jon M. Jenkins}
\affiliation{NASA Ames Research Center, Moffett Field, CA 94035, USA}

\author[0000-0002-3481-9052]{Keivan Stassun}
\affiliation{Department of Physics and Astronomy, Vanderbilt University, Nashville, TN 37235, USA}

\author[0000-0002-6778-7552]{Joseph D. Twicken}
\affiliation{SETI Institute, Mountain View, CA  94043, USA}
\affiliation{NASA Ames Research Center, Moffett Field, CA  94035, USA}

\author[0000-0001-6763-6562]{Roland Vanderspek}
\affiliation{Department of Physics and Kavli Institute for Astrophysics and Space Research, Massachusetts Institute of Technology, Cambridge, MA 02139, USA}

\author[0000-0002-4265-047X]{Joshua N. Winn}
\affiliation{Department of Astrophysical Sciences, Princeton University, Princeton, NJ 08544, USA}


\begin{abstract}
During its 2-year Prime Mission, \tess\ observed over 232,000 stars at a 2-min cadence across $\sim$70\% of the sky.  These data provide a record of photometric variability across a range of astrophysically interesting time scales, probing stellar rotation, stellar binarity, and pulsations.  We have analyzed the \tess\ 2-min light curves to identify periodic variability on timescales 0.01--13\,days, and explored the results across various stellar properties.  We have identified over 46,000 periodic variables with high confidence, and another 38,000 with moderate confidence.  These light curves show differences in variability type across the HR diagram, with distinct groupings of rotational, eclipsing, and pulsational variables.  We also see interesting patterns across period-luminosity space, with clear correlations between period and luminosity for high-mass pulsators, evolved stars, and contact binary systems, a discontinuity corresponding to the Kraft break, and a lower occurrence of periodic variability in main-sequence stars on timescales of 1.5 to 2 days. The variable stars identified in this work are cross-identified with several other variability catalogs, from which we find good agreement between the measured periods of variability. There are $\sim$65,000\,variable stars that are newly identified in this work, which includes rotation rates of low-mass stars, high-frequency pulsation periods for high-mass stars, and a variety of giant star variability.
\end{abstract}



\section{Introduction} \label{sec:intro}
Changes in the observed flux of a star over time, generally referred to as stellar photometric variability, can be attributed to a range of astrophysical processes. Stellar variability may be stochastic in nature, such as rapid changes in flux caused by flares, mass ejections, or novae, but changes in flux may also occur periodically. Periodic (or semi-periodic) variations in stellar flux have been linked to pulsations, rotation, eclipses, and other conditions or processes. The details of these variations can provide information about the dynamics, internal structure, and composition of stars, along with fundamental physical properties, such as mass, radius, and age \citep[see][and references therein]{Soderblom10, Chaplin13, Hekker17, Kochukhov21, Kurtz22}. Ages, for example, are typically the most difficult stellar property to measure, but gyrochronology has allowed stellar ages to be constrained from the rotation rates of Sun-like stars since they are known to lose angular momentum and spin down over time \citep{Irwin09, Meibom09, van_Saders13, Angus19}. The rate of angular momentum loss is connected to a star's internal structure and angular momentum loss through stellar wind, such that a star's surface activity can also be used to probe the structure and dynamo processes inside stars \citep{Bohm-Vitense07, Ferreira_Lopes15, Han21}. 

Characterizing the variability of stars is also important for understanding the nature of exoplanetary systems. Exoplanets are typically detected indirectly, such as through the transit of a planet in front of its host star or the radial velocity changes of a star due to the orbit of a planet. Therefore, the determination of the properties of exoplanets relies on our understanding of their host stars. Planetary radii, for example, are a critical measurement for distinguishing between terrestrial planets, giant planets, brown dwarfs, and stellar companions \citep{Kane12, Fulton17, Owen17}. However, uncertainties in stellar radii lead to corresponding uncertainties in planetary radii \citep{Seager07, Ciardi15, Hirsch17, Kane14, Kane18}. Active stars with large flux amplitude variations have also been shown to masquerade as planetary signatures in radial velocity measurements \citep[e.g.,][]{Henry02, Robertson14, Robertson14-1, Robertson15, Kane16, Hojjatpanah20, Prajwal22, Simpson22}, while at the same time also potentially hiding the presence of small planets in their system. Furthermore, stellar variability directly impacts the insolation flux received by exoplanets, with implications for atmospheric erosion \citep{Lammer03, Murray-Clay09, Owen13, Dong17, Kreidberg19, Sakuraba19, Kane20} and habitability \citep{Tarter07, Segura10, Dong18, Howard20b}. This is especially true for planets around M-dwarf stars, which are known to exhibit particularly strong outbursts of ultraviolet and X-ray radiation as part of their activity cycles \citep{Tarter07, Segura10, Jackman19, Gunther20, Howard20b}. High-energy particles and radiation expelled from the star as a result of its magnetic or chromospheric activity can be damaging to the atmospheres of close-in planets and any potential life on the planet \citep{Tarter07, Segura10, Dong18, Gunell18, Howard20b}. Addressing questions about how planetary atmospheres evolve in these systems requires understanding the influence of its host star, including its flux variability. 

Wide-field variability surveys, by design, are able to constrain the fraction of variable stars across their entire field of view down to a certain brightness limit. This brute-force approach typically aims to identify rare transients, such as supernovae, transits of exoplanets in wide orbits, or variable stars along the instability strip, but also enables an understanding of variability across nearly the entire sky. Since a star's variability is linked to its evolutionary state, measuring stellar variability across the sky can be used to understand the evolution history of our galaxy. Extensive work has already been done to detect photometric variability across the sky through ground-based surveys, such as the All-Sky Automated Survey \citep[ASAS;][]{Pojmanski02}, the Optical Gravitational Lensing Experiment \citep[OGLE;][]{Udalski08}, the Kilodegree Extremely Little Telescope \citep[KELT;][]{Oelkers18}, the Asteroid Terrestrial-impact Last Alert System \citep[ALTAS;][]{Heinze18}, the All-Sky Automated Survey for Supernovae \citep[ASAS-SN;][]{Jayasinghe18}, and the Next-Generation Transit Survey \citep[NGTS;][]{Briegal22}. However, ground-based surveys are limited by seeing conditions, weather, and inability to observe during the day. Furthermore, variability studies from space-based surveys, such as the \kepler\ space telescope \citep{Borucki10} and \textit{Gaia} \citep{Gaia_Collaboration16}, have found that the fraction of variable stars across the sky has been limited by photometric precision rather than astrophysical variability \citep{Ciardi11, Basri11, Gaia_Collaboration19, Briegal22}. The Transiting Exoplanet Survey Satellite (\tess) used a tile-based strategy to gradually sample about 70\% of the sky during its first two years of operation \citep{Ricker14, Ricker15}, making it an ideal instrument for a comprehensive study of stellar variability within the galaxy. Furthermore, the near-infrared bandpass utilized by \tess\ allows for high precision photometry of M dwarf stars in the Solar neighborhood.

In this paper, we present a stellar variability catalog based on our search for photometric variability on timescales shorter than 13\,days across nearly the entire sky using the 2-min cadence photometry obtained during the \tess\ Prime Mission. In \autoref{sec:methods} we outline the preparation of the \tess\ photometry, the Fourier analysis of the data, special considerations as to how the \tess\ mission design affects our study of variability, and the methodology used to identify variable signatures that are astrophysical in nature---as opposed to being caused by spacecraft systematics. Our stellar variability catalog, constructed to have high reliability at the expense of high completeness, is presented in \autoref{sec:results}. A review of the demographics and population statistics of the stars in the variability catalog are presented in \autoref{sec:discussion}, alongside a comparison to previous works and our planned follow-up investigations. Finally, we summarize this work in \autoref{sec:summary}.
%
%
%


\section{Data and Methods}\label{sec:methods}

\subsection{TESS Photometry} \label{sec:data}
\begin{figure*}
\plottwo{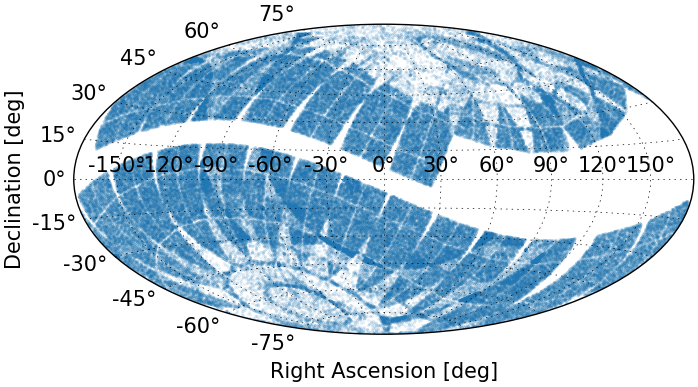}{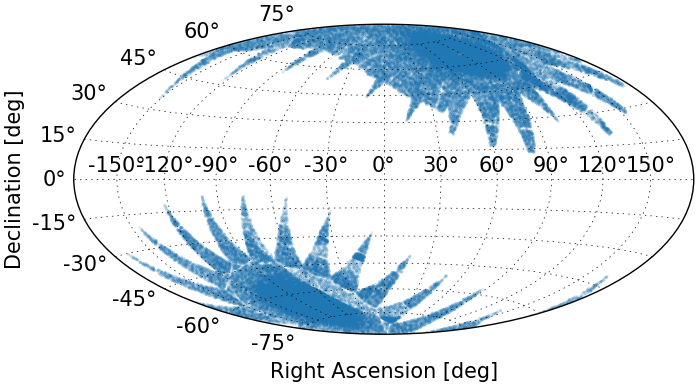}
\caption{Distribution in equatorial coordinates of stars observed at 2-min cadence by \tess\ during its first two years of operations: the \tess\ Prime Mission. \textit{Left:} Stars that were observed in only one sector of \tess\ observations (27.4\,days). \textit{Right:} Stars that were observed in two or more sectors of \tess\ observations ($>$54.8\,days).}
\label{fig:skyDist}
\end{figure*}

Here we briefly summarize information about the \tess\ mission that is relevant to this work, and refer the reader to \citet{Ricker14, Ricker15} for more detail. The \tess\ spacecraft is in a 13.7\,day orbit around the Earth in a 2:1 resonance with the Moon and observes a $24^\circ\times96^\circ$ region of the sky using four optical CCD cameras that are sensitive to light in the range of 600--1000\,nm. Time-series photometry is obtained in 27.4\,day segments, known as sectors, after which the spacecraft repoints to observe a different segment of the sky, maintaining a general pointing anti-Sun. During the \tess\ Prime Mission (2018~July~25--2020~July~04), $\sim$200,000 targets distributed across $\sim$70\% of the sky were observed at 2-min cadence for at least the duration of one sector of observations. Targets located closer to the ecliptic poles were more likely to be observed in multiple sectors, at most being observed nearly continuously for 1\,year (13\,sectors). \autoref{fig:skyDist} shows the distribution of the \nSingle\ stars observed by a single \tess\ sector compared to the \nMult\ stars observed in two or more sectors. Information about the \tess\ targets is provided in the \tess\ Input Catalog \citep[TICv8;][]{Stassun18,Stassun19}, which provides both observational and astrophysical information about each target star and those in the surrounding star field. Since the \tess\ pixels are 21\,arcsec and are subject to blending between neighboring stars, the TICv8 includes a crude calculation of the degree of flux contamination for each star, called the contamination ratio (\texttt{CONTRATIO}). The \tess\ Input Catalog (\dataset[DOI: 10.17909/fwdt-2x66]{http://dx.doi.org/10.17909/fwdt-2x66}), full-frame images (\dataset[DOI: 10.17909/3y7c-wa45]{http://dx.doi.org/10.17909/3y7c-wa45}), time-series photometry (\dataset[DOI: 10.17909/t9-nmc8-f686]{http://dx.doi.org/10.17909/t9-nmc8-f686}) are publicly available through the Mikulski Archive for Space Telescopes\footnote{\url{https://archive.stsci.edu/}} (MAST). 

While the \tess\ Prime Mission (Sectors 1--26) observed a total of \nAll\ unique stars in 2-min cadence ($\sim$20,000\,stars observed per sector), we restrict our periodicity search to the \nSamp\ stars that are brighter than $T=14$ and are not severely blended with neighboring stars (\texttt{CONTRATIO} $<$ 0.2). We also only examine stars with physical parameters listed in the TIC, which does not include white dwarfs and many subdwarfs. We use the 2-min cadence time-series photometry that were processed by the Science Processing Operations Center (SPOC) pipeline \citep{Jenkins16}, known as the Presearch Data Conditioning Simple Aperture Photometry \citep[PDCSAP;][]{Stumpe12, Stumpe14, Smith12} light curves. The PDCSAP light curves are based on an optimal aperture extraction of the raw photometry that was detrended using co-trending basis vectors in order to correct for common instrument systematics. We apply additional post-processing to the light curves, including removing data that are flagged as being anomalous in quality, trimming any known and candidate planetary transit events, and removing outliers that are $>$5$\sigma$ from the RMS of the light curve. Transit events are removed using the reported orbital period and a transit duration that is 10\% longer than that which is reported in the \tess\ Object of Interest (TOI) catalog\footnote{Accessed 2020 December 13.} \citep{Guerrero21}. In addition to removing data that has been flagged as being anomalous quality by the pipeline, we remove observations between 1347.0--1349.8\,TJD during which significant flux scatter is observed in most Sector\,1 light curves due to a known spacecraft stability issue that is discussed in the \tess\ data release notes. This section of Sector\,1 was the only set of data that we removed due to pervasive problems across many light curves.
%
%
%


\subsection{Periodicity Search Tools} \label{sec:periodogram}
Stellar variability manifests in many forms, and can be periodic, semiperiodic, or stochastic. In this paper, we mostly concern ourselves with continuous periodic sinusoidal-like photometric variability. Such behavior is associated with stellar pulsations, rotational modulations due to starspots, and ellipsoidal variations in stellar binaries and can be detected with the Lomb-Scargle (LS) periodogram \citep{Lomb76, Scargle82}.

We compute the fast LS periodogram using the \texttt{astropy.timeseries.LombScargle} function and sampling at equally spaced frequencies that are 1.35\,min$^{-1}$ apart, deliberately slightly oversampling the time series for maximal frequency resolution. The highest power peak in the periodogram is used to determine the most significant periodic signature of the light curve. The frequency, uncertainty, and significance (i.e., normalized power) of the most significant periodic signature are measured using a quadratic function fit to the highest power peak in the LS periodogram. The inverse of the frequency with the highest periodogram power is taken to represent the stellar variability period ($P=1/f$). A sinusoidal function is fit to the light curve and defined by
\begin{eqnarray}
F(t) &= A\cos\left[{\frac{2\pi}{P}(t-t_{0})}\right]
\mathrm{,}
\label{eq:sinewave}
\end{eqnarray}
where $A$ is the amplitude of the sine wave, $t-t_0$ represents the time since the first maximum relative to each point in time, and $P$ is the period with the highest power from the LS periodogram. The reduced chi-squared, $\chi^2_\nu$, is used to determine the goodness-of-fit of the sinusoidal function. 

In order to distinguish continuous periodic variability from more punctuated periodic variability, such as planetary transits and stellar eclipses, we also utilize the auto-correlation function (ACF). The ACF requires data that are equally spaced in time, therefore we interpolate any missing observations in time and set their relative fluxes to zero. The ACF is measured by correlating the interpolated data with itself and then smoothing the result using a Gaussian kernel. The lag time of the ACF that has the highest correlation (i.e., strongest peak) is fit with a quadratic function to identify the variability period, its uncertainty, and its significance (i.e., normalized power). To determine the goodness-of-fit for the ACF, the light curve is binned into 100 equally spaced points in phase space and a cubic interpolation of the binned points is used to represent the ACF model when calculating the $\chi_\nu^2$ of the ACF model compared to the light curve data. 

\subsection{Considerations of the \tess\ Photometry} \label{sec:consideration}
There are several considerations that affect our periodogram search in accordance to the nature of the \tess\ photometry. In particular, continuous segments of observations take place over a single spacecraft orbit, the placement of each target on the CCD cameras changes between \tess\ sectors, and the spacecraft performs periodic adjustments in pointing that can introduce systematic noise into the photometry. 

Continuous \tess\ observations occur during the majority of each spacecraft orbit, but there is a $\sim$1\,day pause in observations when the spacecraft approaches perigee in order to downlink the data. Otherwise, observing conditions remain consistent over the course of two spacecraft orbits, which makes up the duration of a \tess\ sector ($\sim$27\,days). Every \tess\ target is observed nearly continuously for at least one sector, such that we can search for repeatable signatures in the \tess\ light curves up to $\sim$13\,days. However, periodic signatures that are measured to be closer to the 13-day limit are subject to greater uncertainties than those on shorter timescales. 

There are \nMultSamp\ stars in our sample that have been observed in more than one sector of \tess\ photometry. Due to the observational design of the \tess\ mission, each \tess\ sector observes a different region of the sky such that any given star is not observed on the same part of the detector between sectors. The differing positions of the star on the detector introduces systematics into the time-series photometry that are typically not adequately removed with a simple normalization offset. Therefore, we choose to defer an investigation of long-period variability ($>$13\,days) to a future work (Fetherolf et al. in prep.).  In this work, we instead use these stars to understand the reliability of detecting stars that are astrophysically variable---as opposed to being caused by spacecraft systematics---in our catalog by ensuring that the same periodic signal for a given star can be recovered in different sectors (see \autoref{sec:reliability}). 

\begin{figure*}
\plotone{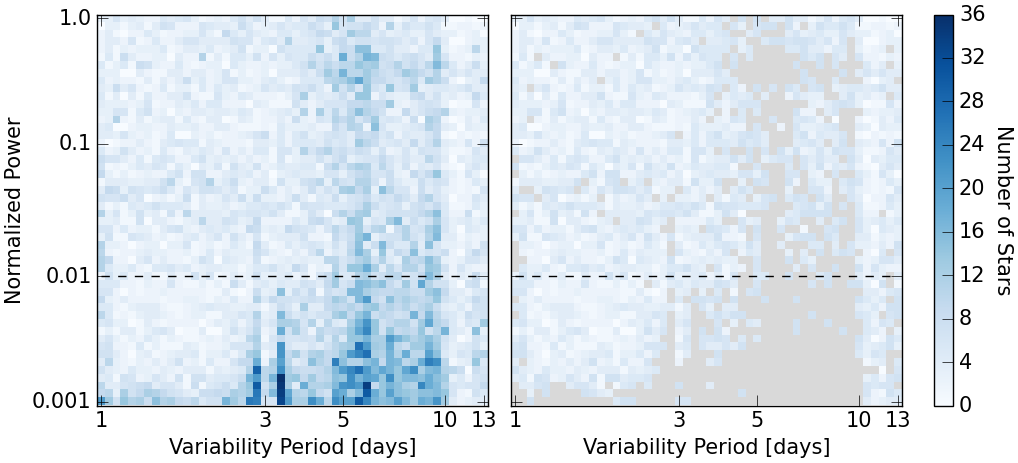}
\plotone{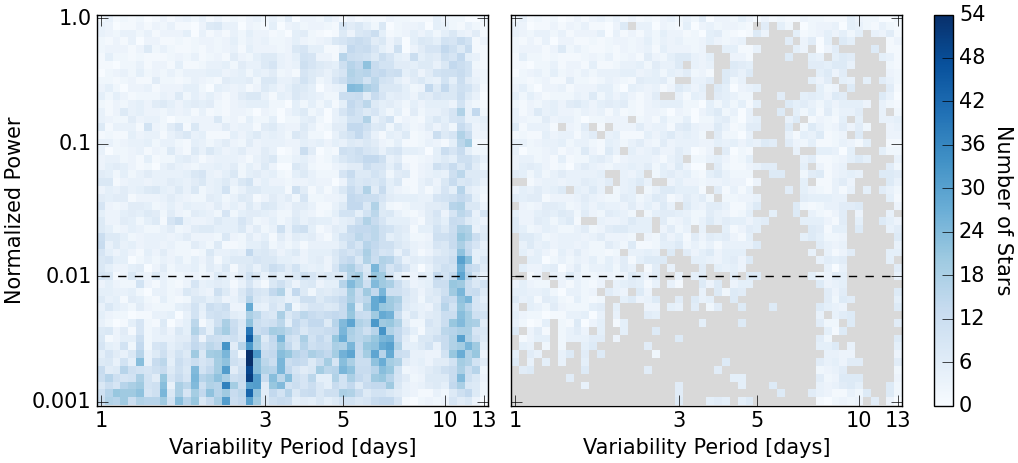}
\caption{Normalized LS power versus the most significant photometric periodicity detected from the light curves for stars that were observed during \tess\ Sectors 5 (\textit{top}) and 17 (\textit{bottom}). The left panels show the density of stars in power-period space. The right panels show in gray which bins were deemed high density ($>$7\,stars), such that the stars were removed from the catalog. Stars that fall in high density regions of power-period space are likely exhibiting periodic modulations in their light curves that can be attributed to instrument systematics. The horizontal dashed line shows the normalized LS power cut that is applied to the final catalog (Tables~\ref{tab:1peak}--\ref{tab:ACF}).}
\label{fig:density}
\end{figure*}

The spacecraft reaction wheels have their speeds reset (also known as ``momentum dumps'') as frequently as every few days, which can cause periodic changes in the light curve that can masquerade as stellar variability. These systematic variations are not consistent between individual sectors (becoming less frequent over the mission duration), but they can occasionally be detected in the light curve at high Lomb-Scargle power in some cases. The left panels of \autoref{fig:density} show how the variability periods detected from the single sinusoidal model at 1--13\,days are distributed in normalized power and period space for stars observed during \tess\ Sectors 5 and 17. There are several regions in the normalized power versus period space that contain a higher density of stars, but these periodicities would align between different \tess\ sectors if they were astrophysical in nature. In the interest of developing a stellar variability catalog of high confidence, we elect to remove any stars that fall in high-density regions in power versus period space for measured variability periods $>$1\,day in each \tess\ sector at the cost of removing a few real variable stars from the catalog (see \autoref{sec:search} and right panels of \autoref{fig:density}).\footnote{High-density patterns in power-period space that could be attributed to spacecraft systematics are not observed for variability periods shorter than 1\,day.}
%
%
%


\subsection{Variability Search} \label{sec:search}
\begin{figure*}
\plotone{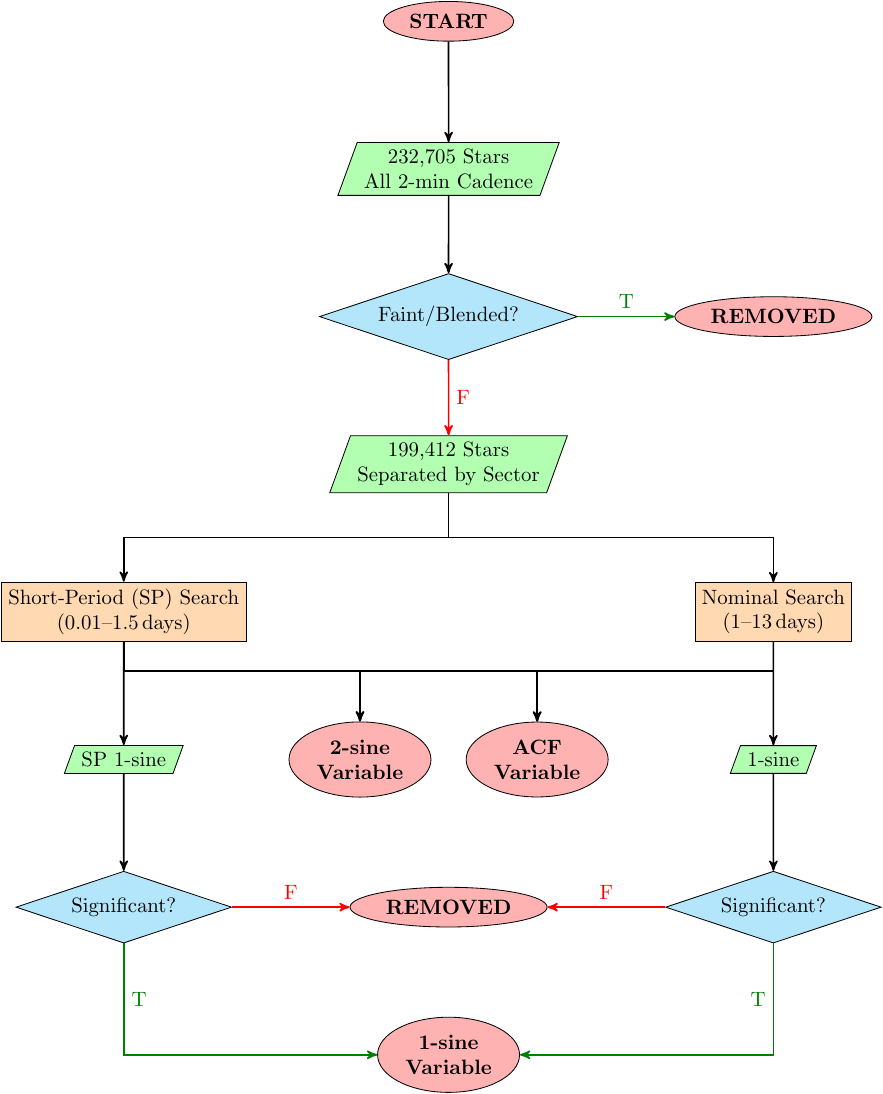}
\caption{A flowchart that summarizes our variability search algorithm, with the direction of flow being indicated by black (always), green (only when True), or red (only when False) arrows. Starting from the stars observed during the \tess\ Prime Mission, we first remove stars that are faint ($T_{\mathrm{mag}}>14$\,mag) or subject to significant blending (\texttt{CONTRATIO} $>$ 0.2). We then search for periodic signatures at 0.01--1.5 and 1--13\,days in each individual sector of \tess\ photometry. Several tests are preformed to determine significance of variability that is best fit with a single sinusoidal model (see \autoref{sec:search}). Stars that make it into one of the ``Variable'' categories are included in Tables~\ref{tab:1peak}--\ref{tab:ACF}.}
\label{fig:flowchart}
\end{figure*}
We perform a periodogram search on \nSamp\ unique stars, which comprises approximately half a million light curves that were obtained in \tess\ Sectors 1--26. Our periodogram search includes two distinct period ranges at 0.01--1.5\,days and 1--13\,days. \autoref{fig:flowchart} shows the overall flow of our periodogram search, where stars that we identify as significantly variable are best characterized by either a single-sinusoidal function (1-sine Variable), a double-sinusoidal function (2-sine Variable), or the ACF (ACF Variable). 

Our default model for the light curve is a single sinusoidal fit, but a double-sinusoidal function fit is also performed when the second-highest peak in the LS periodogram has a normalized power greater than 0.1. The double-sinusoidal fit is accepted as a better model than the single-sinusoidal fit to the light curve variations if there is at least a 25\% improvement in the $\chi_\nu^2$. The ACF fit is performed when both the single- and double-sinusoidal models are poorly fit ($\chi_\nu^2>100$) to the light curve. The ACF is accepted as a better model if the strongest correlation value is greater than 0.5 and the $\chi_\nu^2$ improves compared to the sinusoidal models. A star that is best fit with either a double-sinusoidal model or the ACF is considered significantly variable by default, the purity of which is further explored in \autoref{sec:reliability}. The values for the normalized power cutoffs and $\chi_\nu^2$ improvement were determined through visual inspection of the light curve fits for a preliminary sample of stars.

If a star that passes the initial periodogram search and is not identified as significantly variable with a double-sinusoidal function or the ACF, then it is subject to further vetting after assuming the single-sinusoidal fit to be the best model. First, a separate short-period variability search (0.01--1.5\,days) is performed and used to identify significant variability on short timescales ($<$1.1\,days) when the normalized power is $>$0.01 normalized LS power. Several vetting steps are then applied to remove systematic or long-period ($>$13\,days) variability from the nominal periodogram search at 1--13\,days, with the first being a minimum threshold normalized LS power of 0.01.\footnote{\autoref{fig:density} shows how variability below this threshold can often be attributed to spacecraft systematics.} To retain high purity in the variability catalog, we remove stars that fall in high-density regions in power-period space for variability periods measured to be greater than 1\,day (see \autoref{sec:consideration} and \autoref{fig:density}), where variability is likely due to spacecraft systematics. High-density regions are determined by binning the log power and log period into 50 equally spaced bins (power $=$ 0.001--1.0; $P=$ 1--13\,days), and stars whose variability properties fall within bins that contain seven or more total stars are considered to be spuriously variable---although we reiterate that some real variable stars will be removed during this step. \autoref{fig:density} shows examples from two sectors of the power-period distribution patterns in which stars are removed from our catalog based on the high density in power-period space for Sectors 5 and 17, where different patterns in power-period space can be seen. However, most of the stars that are inevitably removed from the catalog based on the high-density regions in power-period space tend to fall below our minimum normalized LS power threshold of 0.01. Stars that are variable on timescales longer than that which can be properly constrained by the \tess\ single-sector photometry ($\gg$13\,days), or are indistinguishable from uncorrected systematics on these timescales, are identified via two methods and removed from our variability catalog: 1) Stars that exhibit a maximum power in their LS periodogram at the upper limit of the period range searched (i.e., 13\,days) are assumed to exhibit variability at $\gtrsim$13\,days, which is beyond our sensitivity limit for the variability period. 2) We perform a linear fit to each light curve; if the $\chi_\nu^2$ of the linear fit is better than that of the sinusoidal or ACF models, we consider the star to have a long-term trend. Overall, stars that exhibit single-sinusoidal variability are considered significantly variable and, thus, are retained in our variability catalog if they are not caught by the filters for being spuriously variable and if they do not exhibit long-term trends. 

Stars that were observed in multiple \tess\ sectors may be detected as significantly variable in more than one sector. However, the variability period may not match between all sectors. It may be that the detected periodicity in one sector is half or twice that of another sector, the photometric variability for a given sector may be dominated by periodic momentum dumps of the spacecraft, or the star may be exhibiting variability that changes in period or amplitude over time. For the variability catalog reported in this work, we provide only the results from a single sector of \tess\ photometry. The \tess\ sector from which we report the periodogram results for a given star is based on first prioritizing any variability that is best fit by an ACF or double-sinusoidal model, then choosing the periodogram result with the highest normalized power. 
%
%
%



\begin{deluxetable*}{rcrcrrrrrccc}
\tablecaption{Light Curve Fitting and Stellar Properties of Single-Sinusoidal Variables\label{tab:1peak}}
\tablewidth{700pt}
\tabletypesize{\scriptsize}
\tablehead{
\colhead{TIC} & \colhead{Sector} & 
\colhead{$P_\mathrm{var}$} &
\colhead{power} & \colhead{$A$} &
\colhead{$T_0$} & \colhead{$\chi^2_\nu$} &
\colhead{RMS} & \colhead{$T_\mathrm{mag}$\tablenotemark{a}} &
\colhead{$T_\mathrm{eff}$\tablenotemark{a}} & \colhead{$R_*$\tablenotemark{a}} & 
\colhead{$L_*$\tablenotemark{b}} \\ [-0.3cm]
\colhead{~} & \colhead{~} & \colhead{(days)} & \colhead{~} &
\colhead{(ppm)} & \colhead{(BTJD)} & \colhead{~} & \colhead{(ppm)} &
\colhead{~} & \colhead{(K)} & \colhead{(\Rsun)} & \colhead{(\Lsun)}
} 
\startdata
355235442 & 9      & $1.225 \pm 0.024$  & 0.999 & 43096   & 1543.778 & 5.65     & 30751           & 10.15 & 5266  & 1.02  & 0.72   \\
361948797 & 16     & $9.485 \pm 1.361$  & 0.999 & 52113   & 1743.029 & 6.89     & 37109           & 8.77  & 14618 & ---   & ---    \\
285651536 & 12     & $12.365 \pm 2.394$ & 0.998 & 22679   & 1625.519 & 6.07     & 16823           & 7.85  & ---   & ---   & ---    \\
29953651  & 9      & $1.472 \pm 0.033$  & 0.995 & 20919   & 1544.337 & 11.92    & 14845           & 8.10  & ---   & ---   & ---    \\
441807438 & 21     & $1.135 \pm 0.018$  & 0.995 & 33715   & 1871.116 & 19.43    & 23917           & 8.81  & 5181  & 1.22  & 0.97   \\
127256815 & 7      & $1.313 \pm 0.027$  & 0.994 & 26086   & 1491.655 & 9.46     & 18674           & 8.95  & 13833 & ---   & ---    \\
47985275  & 6      & $1.111 \pm 0.021$  & 0.993 & 16295   & 1465.957 & 4.15     & 11605           & 9.37  & 12081 & ---   & ---    \\
303860976 & 11     & $1.275 \pm 0.025$  & 0.993 & 6953    & 1596.913 & 2.62     & 4993            & 7.97  & 10237 & 3.83  & 144.53 \\
117765777 & 6      & $1.101 \pm 0.021$  & 0.992 & 31558   & 1465.491 & 78.11    & 22488           & 7.91  & 11345 & 3.00  & 134.18 \\
401481773 & 10     & $8.580 \pm 1.209$  & 0.992 & 24077   & 1577.481 & 9.97     & 17740           & 8.63  & ---   & ---   & ---    
\enddata
\tablecomments{The 10 stars with the highest LS normalized power are listed here as a reference, but the complete table with several additional columns of information} is available in the machine-readable format.
\tablenotetext{a}{Taken from the TICv8 catalog \citep{Stassun18,Stassun19}.}
\tablenotetext{b}{Calculated from $T_\mathrm{eff}$ and $R_*$.}
\end{deluxetable*}

\begin{deluxetable*}{rcrcrrrrrccc}
\tablecaption{Light Curve Fitting and Stellar Properties of Double-Sinusoidal Variables\label{tab:2peak}}
\tablewidth{700pt}
\tabletypesize{\scriptsize}
\tablehead{
\colhead{TIC} & \colhead{Sector} & 
\colhead{$P_\mathrm{var}$} &
\colhead{power} & \colhead{$A$} &
\colhead{$T_0$} & \colhead{$\chi^2_\nu$} &
\colhead{RMS} & \colhead{$T_\mathrm{mag}$\tablenotemark{a}} &
\colhead{$T_\mathrm{eff}$\tablenotemark{a}} & \colhead{$R_*$\tablenotemark{a}} & 
\colhead{$L_*$\tablenotemark{b}} \\ [-0.3cm]
\colhead{~} & \colhead{~} & \colhead{(days)} & \colhead{~} &
\colhead{(ppm)} & \colhead{(BTJD)} & \colhead{~} & \colhead{(ppm)} &
\colhead{~} & \colhead{(K)} & \colhead{(\Rsun)} & \colhead{(\Lsun)}
} 
\startdata
56980355  & 8      & $4.490 \pm 0.281$  & 0.984     & 35938       & 1520.424   & 16.25        & 25262               & 9.08  & 9684  & 4.95  & 193.51  \\
          &        & $6.045 \pm 0.487$  & 0.316     & 3387        & 1518.653   &              &                     &       &       &       &         \\
140797524 & 8      & $9.576 \pm 1.287$  & 0.979     & 87883       & 1525.411   & 904.94       & 61352               & 7.24  & 4691  & 8.67  & 32.69   \\
          &        & $6.216 \pm 0.516$  & 0.332     & 7859        & 1519.354   &              &                     &       &       &       &         \\
455000299 & 11     & $7.911 \pm 0.899$  & 0.977     & 30308       & 1596.905   & 7.18         & 21797               & 10.09 & 4397  & 1.18  & 0.47    \\
          &        & $5.220 \pm 0.398$  & 0.417     & 3457        & 1598.956   &              &                     &       &       &       &         \\
33298241  & 7      & $11.784 \pm 2.262$ & 0.968     & 47926       & 1499.506   & 86.60        & 34356               & 8.88  & 10636 & 4.02  & 185.68  \\
          &        & $6.944 \pm 0.546$  & 0.143     & 4868        & 1491.729   &              &                     &       &       &       &         \\
291366757 & 12     & $8.328 \pm 1.009$  & 0.965     & 30518       & 1632.860   & 178.56       & 21593               & 7.18  & 8231  & 3.56  & 52.26   \\
          &        & $5.580 \pm 0.379$  & 0.210     & 3800        & 1628.350   &              &                     &       &       &       &         \\
255585244 & 7      & $11.819 \pm 2.360$ & 0.965     & 36199       & 1502.757   & 5.06         & 25681               & 9.67  & ---   & ---   & ---     \\
          &        & $6.940 \pm 0.550$  & 0.210     & 5036        & 1492.010   &              &                     &       &       &       &         \\
202414513 & 17     & $12.515 \pm 2.167$ & 0.964     & 16514       & 1769.312   & 9.51         & 10897               & 8.53  & 3550  & 95.10 & 1289.69 \\
          &        & $6.958 \pm 0.656$  & 0.449     & 1417        & 1771.497   &              &                     &       &       &       &         \\
258908663 & 11     & $6.147 \pm 0.560$  & 0.957     & 52026       & 1600.852   & 6.22         & 36753               & 11.62 & 3731  & 0.88  & 0.13    \\
          &        & $10.151 \pm 1.458$ & 0.383     & 11196       & 1600.146   &              &                     &       &       &       &         \\
57065604  & 9      & $11.338 \pm 2.502$ & 0.953     & 57539       & 1550.353   & 547.86       & 41710               & 7.99  & 11179 & 2.52  & 88.87   \\
          &        & $6.693 \pm 0.524$  & 0.273     & 7994        & 1549.570   &              &                     &       &       &       &         \\
280161519 & 14     & $12.859 \pm 2.680$ & 0.951     & 12086       & 1689.719   & 14.75        & 8704                & 8.12  & 3672  & 59.03 & 568.85  \\
          &        & $7.703 \pm 0.572$  & 0.154     & 1533        & 1687.899   &              &                     &       &       &       &         
\enddata
\tablecomments{The 10 stars with the highest LS normalized power are listed here as a reference, but the complete table with several additional columns of information} is available in the machine-readable format.
\tablenotetext{a}{Taken from the TICv8 catalog \citep{Stassun18,Stassun19}.}
\tablenotetext{b}{Calculated from $T_\mathrm{eff}$ and $R_*$.}
\end{deluxetable*}

\begin{deluxetable*}{rcrcrrrrrccc}
\tablecaption{Light Curve Fitting and Stellar Properties of ACF Variables\label{tab:ACF}}
\tablewidth{700pt}
\tabletypesize{\scriptsize}
\tablehead{
\colhead{TIC} & \colhead{Sector} & 
\colhead{$P_\mathrm{var}$} &
\colhead{correlation} & \colhead{$A$} &
\colhead{$T_0$} & \colhead{$\chi^2_\nu$} &
\colhead{RMS} & \colhead{$T_\mathrm{mag}$\tablenotemark{a}} &
\colhead{$T_\mathrm{eff}$\tablenotemark{a}} & \colhead{$R_*$\tablenotemark{a}} & 
\colhead{$L_*$\tablenotemark{b}} \\ [-0.3cm]
\colhead{~} & \colhead{~} & \colhead{(days)} & \colhead{~} &
\colhead{(ppm)} & \colhead{(BTJD)} & \colhead{~} & \colhead{(ppm)} &
\colhead{~} & \colhead{(K)} & \colhead{(\Rsun)} & \colhead{(\Lsun)}
} 
\startdata
451949522 & 10     & $0.454 \pm 0.002$ & 0.937 & 105168  & 1569.842 & 5.28     & 36430           & 8.94  & 4962 & 1.77 & 1.70  \\
162432383 & 10     & $0.324 \pm 0.002$ & 0.935 & 12114   & 1569.524 & 70.84    & 4800            & 8.07  & 7308 & 1.38 & 4.86  \\
41170667  & 9      & $0.487 \pm 0.002$ & 0.932 & 200327  & 1543.618 & 14.06    & 69559           & 10.30 & 6853 & 3.51 & 24.41 \\
24662304  & 6      & $0.545 \pm 0.007$ & 0.932 & 13671   & 1465.473 & 829.76   & 5922            & 6.29  & ---  & ---  & ---   \\
285413207 & 7      & $0.638 \pm 0.003$ & 0.931 & 44123   & 1492.088 & 5.75     & 16092           & 9.96  & 6254 & 1.15 & 1.81  \\
43216747  & 22     & $0.438 \pm 0.002$ & 0.930 & 99613   & 1899.351 & 5.43     & 35717           & 10.35 & 5083 & 1.10 & 0.73  \\
132764448 & 7      & $0.418 \pm 0.001$ & 0.930 & 29187   & 1491.636 & 3.41     & 10413           & 9.34  & 4612 & 0.96 & 0.38  \\
157212164 & 7      & $0.572 \pm 0.002$ & 0.929 & 93374   & 1491.733 & 6.98     & 32852           & 10.51 & 5647 & 0.89 & 0.73  \\
424721218 & 20     & $0.388 \pm 0.001$ & 0.929 & 334262  & 1842.409 & 2.55     & 119979          & 11.64 & 7165 & 5.05 & 60.44 \\
140132301 & 9      & $0.329 \pm 0.001$ & 0.928 & 201079  & 1543.415 & 69.62    & 71698           & 9.32  & ---  & ---  & ---   
\enddata
\tablecomments{The 10 stars with the highest correlation value are listed here as a reference, but the complete table with several additional columns of information is available in the machine-readable format.}
\tablenotetext{a}{Taken from the TICv8 catalog \citep{Stassun18,Stassun19}.}
\tablenotetext{b}{Calculated from $T_\mathrm{eff}$ and $R_*$.}
\end{deluxetable*}


\section{Results}\label{sec:results}

\subsection{Variability Catalog} \label{sec:catalog}

\begin{figure*}
\epsscale{1.1}
\plotone{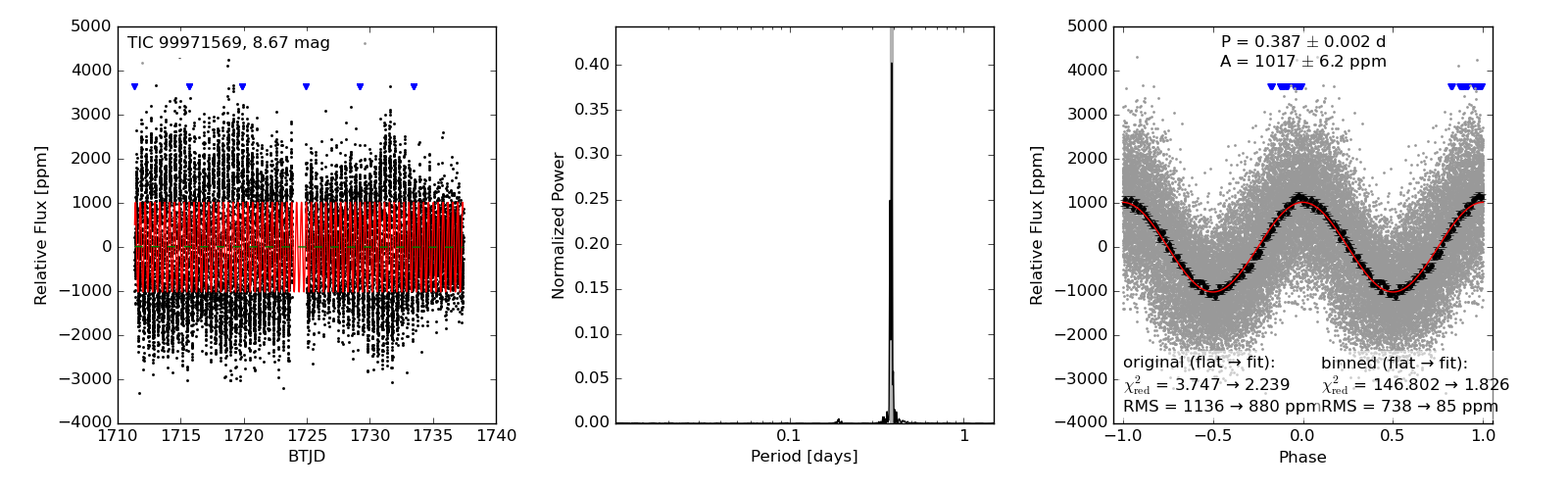}
\plotone{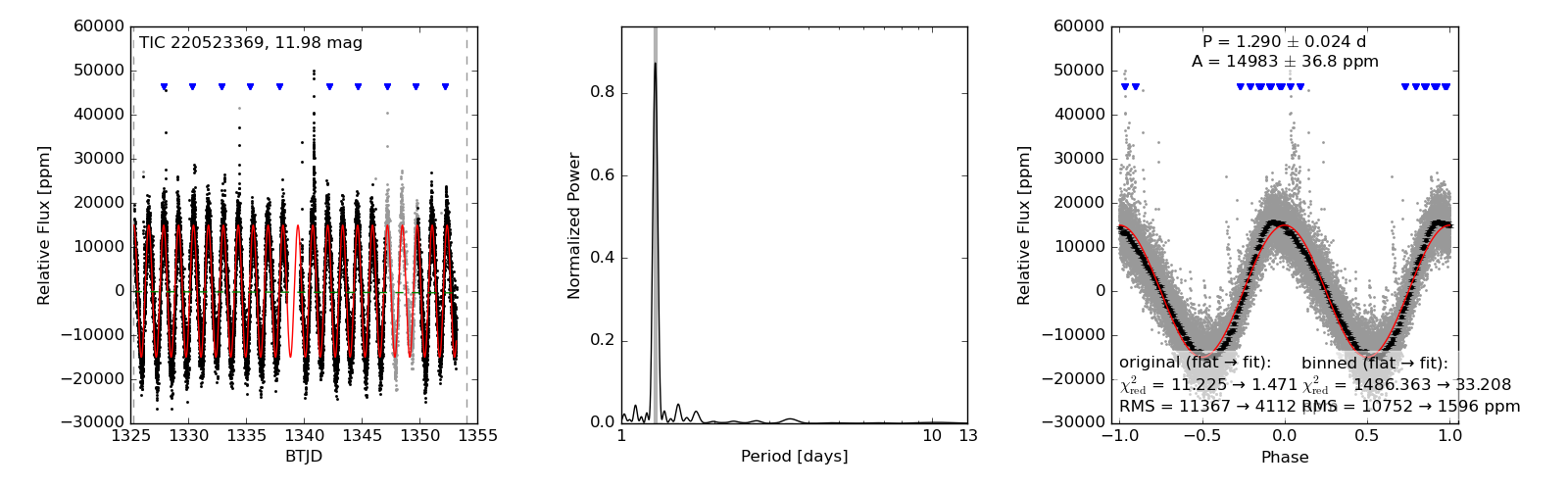}
\plotone{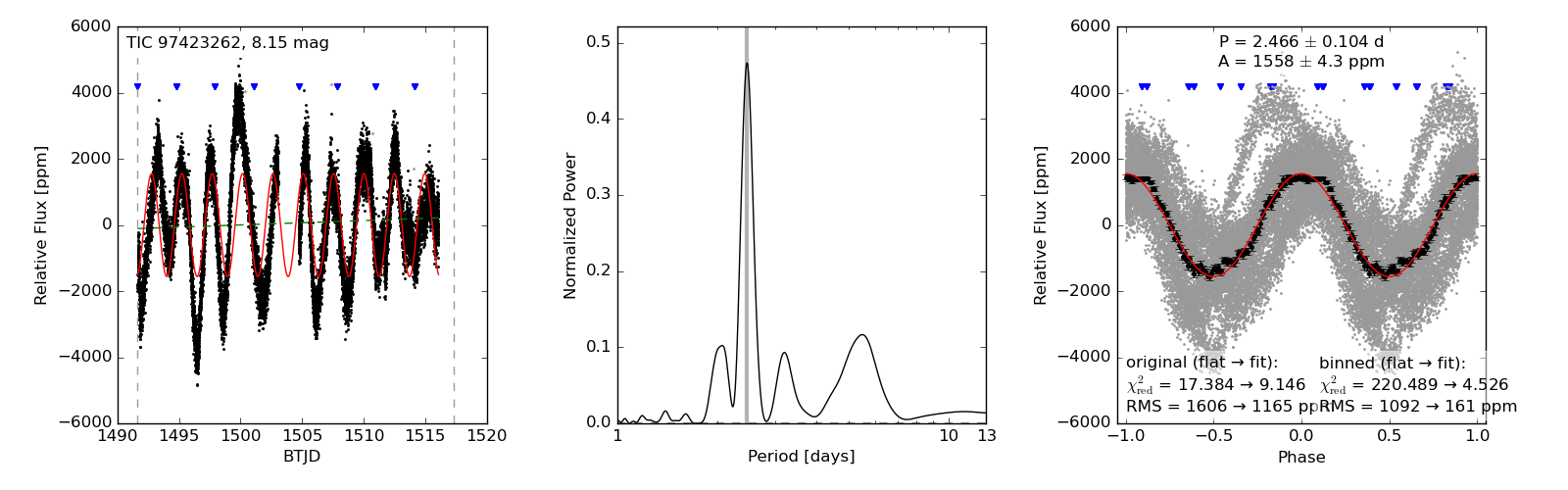}
\caption{Single sector light curve (\textit{left column}), periodogram (\textit{center column}), and phase-folded light curve (\textit{right column}) for 3 examples of stars that are best characterized by a single-sinusoidal function. The black points in the right panels show the medians of 100 bins for the phase curve. The red curve represents the best-fit sinusoidal function and the blue triangles denote spacecraft momentum dump timings. The examples shown include a quickly rotating F star (\textit{top row}), an M dwarf with a fully convective interior (\textit{center row}), and a red giant star (\textit{bottom row}).}
\label{fig:ex_1peak}
\end{figure*}

\begin{figure*}
\epsscale{1.1}
\plotone{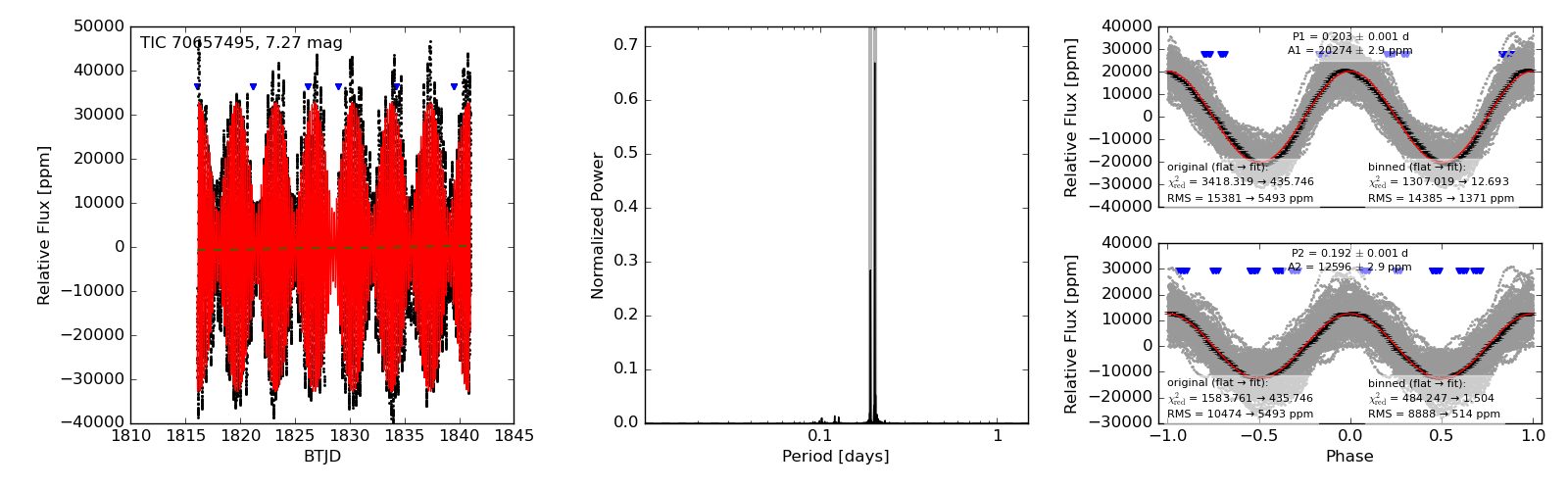}
\plotone{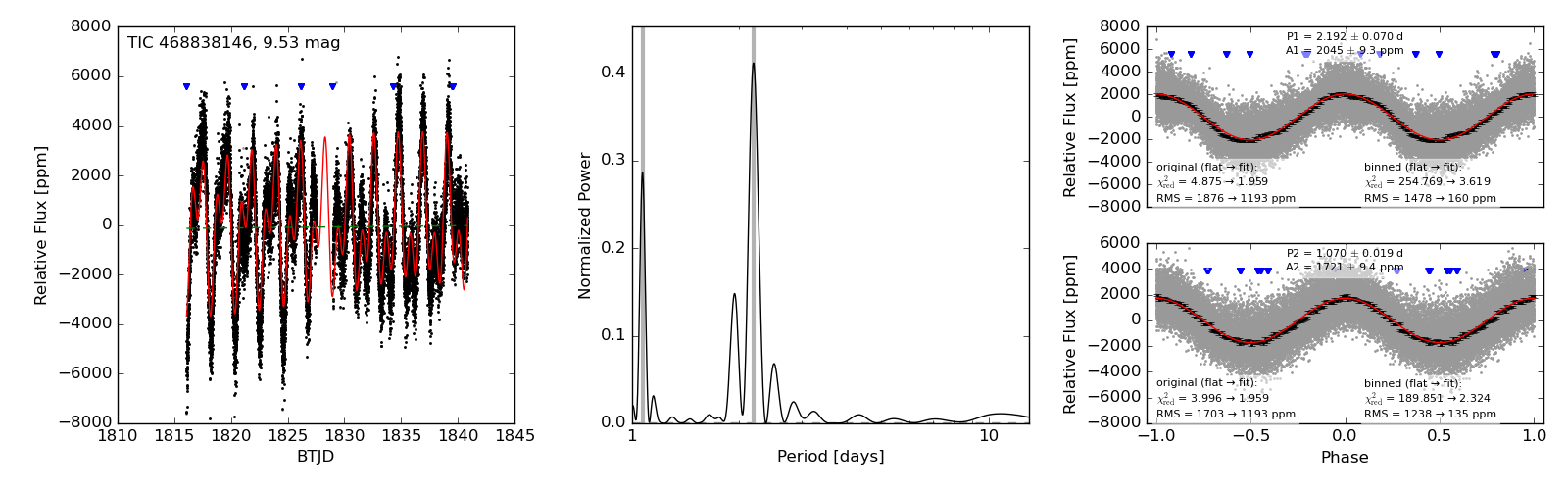}
\plotone{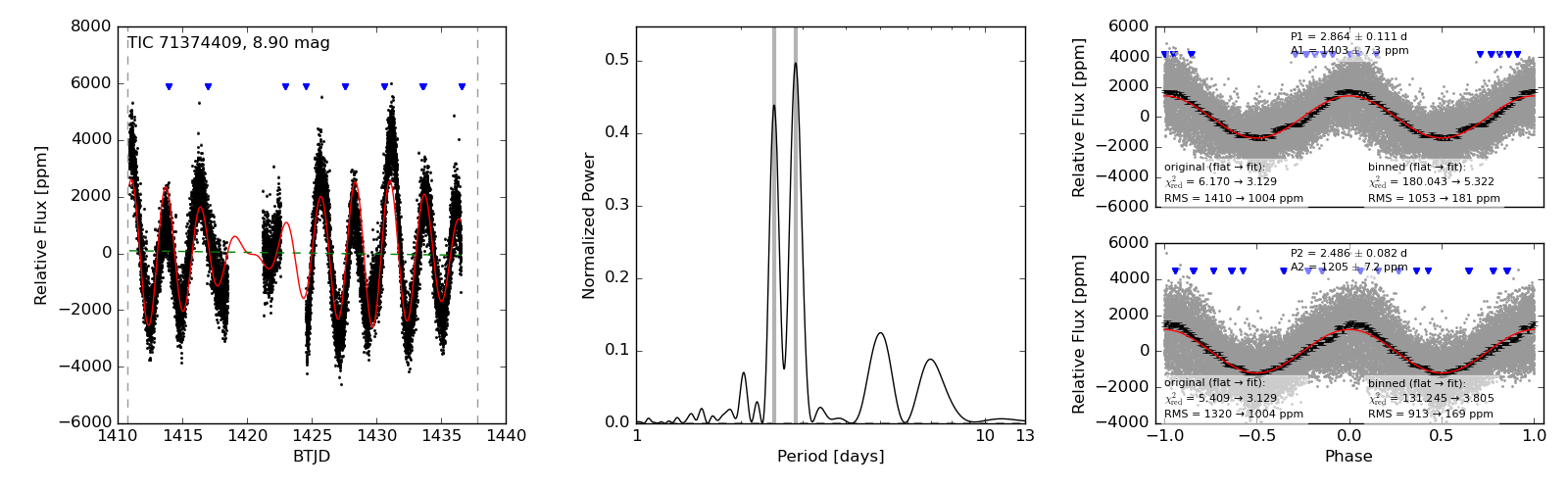}
\caption{Same as \autoref{fig:ex_1peak}, except for stars that are best characterized by a double-sinusoidal function. The right panel shows the phase-folded light curve for the two periodicities that are identified. The examples shown include a $\delta$~Scuti pulsator (\textit{top row}), an F star with a harmonic rotation (\textit{center row}), and a giant star (\textit{bottom row}).}
\label{fig:ex_2peak}
\end{figure*}

\begin{figure*}
\epsscale{1.1}
\plotone{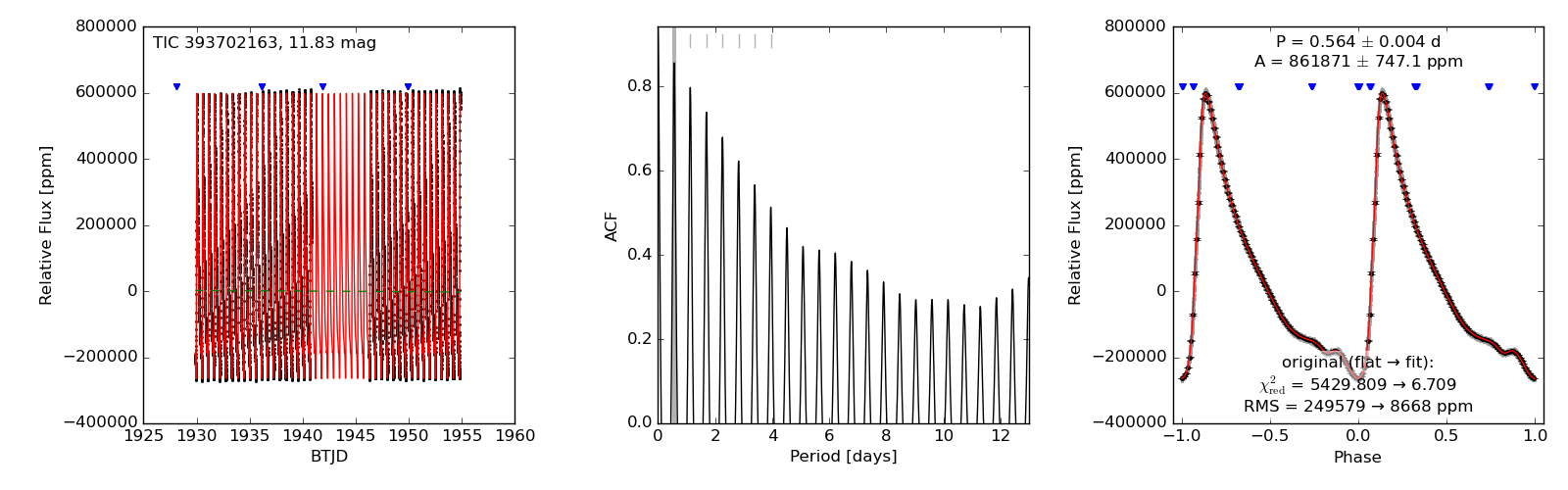}
\plotone{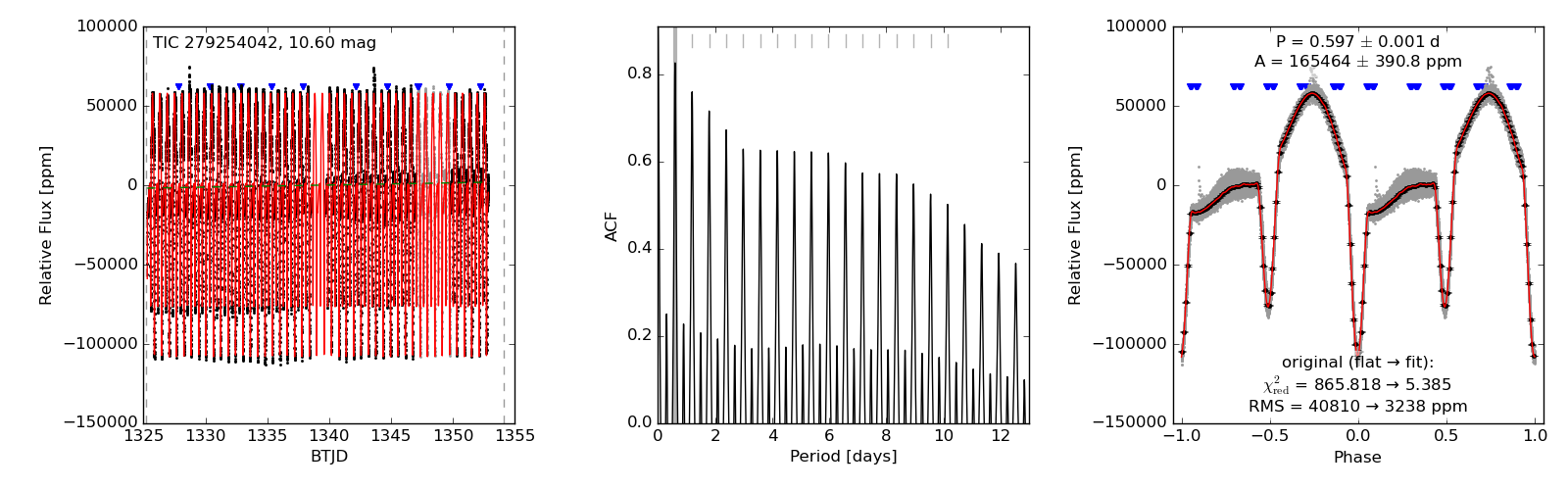}
\plotone{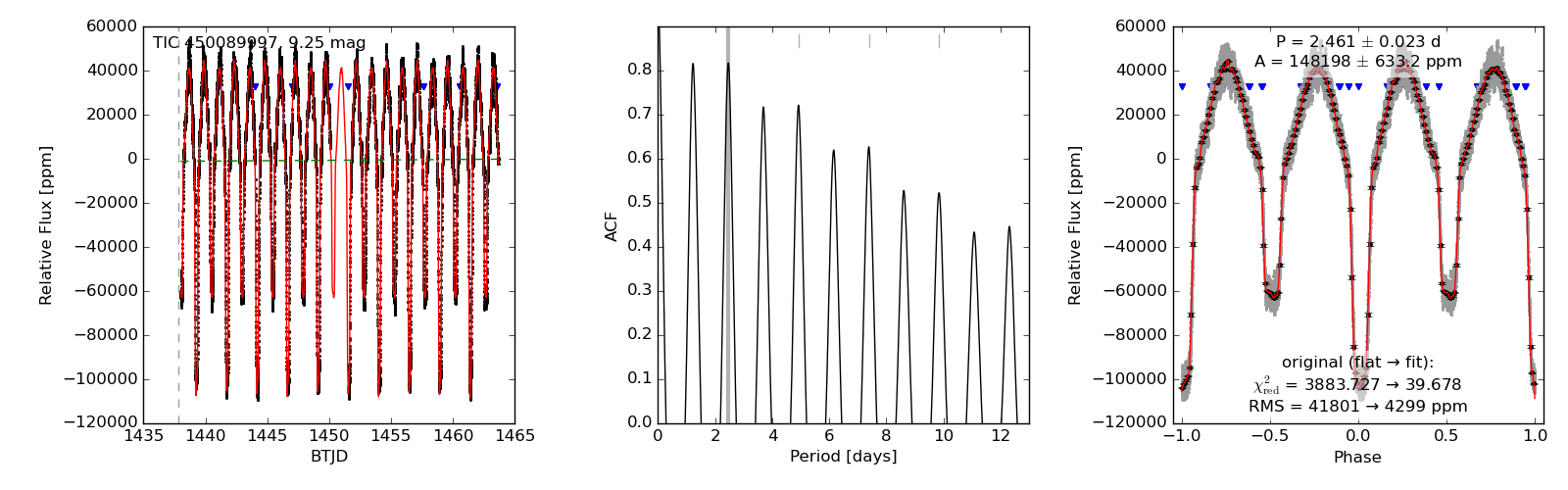}
\caption{Same as \autoref{fig:ex_1peak}, except for stars that are best characterized by an ACF. The center panel shows the correlation over the range of searched variability periods. The red curve represents an interpolation between each median binned point in the phase curve. The examples shown include an RR~Lyrae variable star (\textit{top row}), an eclipsing binary with a V-shaped eclipse (\textit{center row}), and an eclipsing binary with ellipsoidal modulations (\textit{bottom row}).}
\label{fig:ex_ACF}
\end{figure*}

Tables~\ref{tab:1peak}--\ref{tab:ACF} list the \nSine\ stars that are significantly variable when described by a simple sinusoidal function (\nSineSig\ stars with $>$0.1 normalized power), \nDsine\ stars that exhibit double-sinusoidal variability, and \nACF\ stars that are otherwise strictly periodic and best characterized with the ACF, for a total of \nTotal\ unique stars that are included in the variability catalog as a whole. The variability period ($P_\mathrm{var}$) and normalized power (or correlation) are determined from the LS periodogram (or ACF). The amplitude ($A$) and time of maximum flux ($T_0$) of the light curve modulations---and the corresponding reduced chi-squared statistic ($\chi_\nu^2$)---are measured from the best-fit single- or double-sinusoidal function. The \tess\ magnitude ($T_\mathrm{mag}$), stellar effective temperature ($T_\mathrm{eff}$), and stellar radius ($R_*$) are taken from the TICv8 catalog \citep{Stassun18, Stassun19} when available, and we calculate stellar luminosities ($L_*$) using the reported radii and temperatures. The online version of the machine-readable catalog includes several additional informational columns including uncertainties from the sinusoidal fitting, metrics for goodness-of-fit, and all of the contents of the TICv8 catalog for these stars (including coordinates). \autoref{tab:1peak}, in particular, includes variability that was detected with at least 0.01 normalized power ($\sim$68,000 stars), but we hold the highest confidence in the variability of stars with $>$0.1 normalized power ($\sim$30,000 stars). The figures and results that follow only include variable stars that were detected with at least 0.1 normalized power (\nTotalSig\ stars). Tables~\ref{tab:1peak}--\ref{tab:ACF} are available in the machine-readable format and figures showing the light curve, LS periodogram, and phase-folded light curve (e.g., see Figures~\ref{fig:ex_1peak}--\ref{fig:ex_ACF}) for each star in the variability catalog are available as a High-Level Science Product (HLSP) on MAST: \dataset[DOI: 10.17909/f8pz-vj63]{http://dx.doi.org/10.17909/f8pz-vj63}.\footnote{\url{https://archive.stsci.edu/hlsp/tess-svc}}

Our default model assumes sinusoidal variability, which will generally encapsulate rotational variability. \autoref{fig:ex_1peak} shows three examples of periodic variability that are best characterized by a single-sinusoidal function. Light curve modulations that are characterized by a single periodic signal can represent a broad range of stellar activity, including rotational variations caused by starspot activity, a dominant pulsation, or overcontact binaries. The examples shown in \autoref{fig:ex_1peak} showcase three examples of rotational and/or pulsational modulations, including an F star that is hotter than the Kraft break (top row; TIC~99971569), an M dwarf star that has not spun down over time due to its fully convective interior (center row; TIC~220523369), and a giant star exhibiting more quasi-periodic rotation (bottom row; TIC~97423262). 

Some stars exhibit multiple periodicities in their light curves, thus being better characterized by a double-sinusoidal function. Examples of more complex light curves that are best characterized by a double-sinusoidal function are shown in \autoref{fig:ex_2peak}. Multi-periodic variations in stars could be attributed to pulsations or differential stellar rotation. The light curves shown in \autoref{fig:ex_2peak} show examples of a $\delta$~Scuti pulsator (top row; TIC~70657495), a rotating F star that is cooler than the Kraft break (center row; TIC~468838146), and a giant star with two close periodicities (bottom row; TIC~71374409). 

Light curves that are significantly periodic in nature, but are not sinusoidal in shape, are best characterized using the ACF. Finally, we show examples of light curves that exhibit strictly periodic variability that is not necessarily sinusoidal in shape in \autoref{fig:ex_ACF}, which are best characterized by the ACF. These light curves typically represent short-period ($<$11\,days) eclipsing binary systems with V-shaped eclipses, but also can include strictly periodic non-sinusoidal rotational variability. The examples shown in \autoref{fig:ex_ACF} represent an RR~Lyrae variable star (top row; TIC~393702163), an eclipsing binary with a V-shaped eclipse and strong Doppler boosting variations (center row; TIC~279254042), and a flat-bottomed eclipsing binary with strong ellipsoidal modulations (bottom row; TIC~450089997).

\subsection{Catalog Reliability}\label{sec:reliability}
We can characterize the reliability of our catalog by examining stars observed in more than one sector. We compare the periodogram search results for the stars across individual sectors to see whether the periodicity properties are consistent between sectors. As stated in \autoref{sec:search}, we select a single \tess\ sector for reporting the variability analysis results in our catalog (Tables~\ref{tab:1peak}--\ref{tab:ACF}) that is based on the periodicity with the highest normalized power. This selected sector is used as the reference when comparing variability analyses from multiple \tess\ sectors. 

There are \nSineMult\ stars that are best characterized by our default model (single-sinusoidal function) and were observed in multiple \tess\ sectors. For 13,968 stars, we find that the periodicity measured from the selected sector agrees within 10\% of either: 1) the periodicity detected from any other individual sector of photometry; 2) twice or half the periodicity detected from other sectors of photometry; 3) the average periodicity measured from each of the individual sectors; or 4) the average periodicity after removing the largest outlier when the star was observed in three or more \tess\ sectors. Stars whose periodic signatures are not in agreement between multiple sectors may still be astrophysically variable, but could be affected by differential spot rotation or poor photometric detrending in one of the sectors. Overall, there are 5,150\ stars in our single-sinusoidal variability catalog that were observed in multiple sectors but we cannot confirm their astrophysical nature, suggesting a possible 27\% degree of contamination in the single-sinusoidal variability catalog (\autoref{tab:1peak}) by non-variable stars. The catalog contains a list of stars with periodicities that have $>$0.01 normalized power, but we hold higher confidence that the stars with $>$0.1 normalized power are intrinsically variable since only 12\% of these stars are unconfirmed from their multiple sectors of photometry.

There are \nDsineMult\ stars that are observed in multiple \tess\ sectors and best characterized by a double-sinusoidal function. We find that, for all but 41 stars, at least one of the two periodicities identified from the selected sector variability analysis is within 10\% of one of the two periodicities identified in other sectors characterized by a double-sinusoidal function. There are only 12 stars that have periodicities that are inconsistent with other sectors beyond 15\% period variance, but after visual inspection of their light curves we find that they are all consistent with being intrinsically variable in nature with their inconsistent periodicities between \tess\ sectors being attributed to differential rotation (i.e., several close significant periodicities). Overall, we infer with high confidence that \textit{all stars identified as exhibiting a double-sinusoidal signature in their light curve are truly variable in nature}---even when identified using only a single sector of \tess\ photometry (see \autoref{tab:2peak}). 

There are \nACFmult\ stars that are observed in multiple \tess\ sectors and are periodic, but not necessarily sinusoidal in nature, and are thus best characterized by an ACF. Stars that are best characterized by an ACF in at least one \tess\ sector are typically consistent with other ACF periodicities available in other sectors within 10\% of the period. There are 7 stars that have inconsistent results, but are visually examined and found that they could all equally be described by a double-sinusoidal function, with several significant periodicities that are not necessarily harmonics of each other. Given that these few exceptions are visually confirmed as being significantly variable in nature, we infer that \textit{all stars with light curves that are best fit with an ACF}---including those that were observed in only a single sector of \tess\ photometry---are confirmed as being truly variable in nature with high confidence (see \autoref{tab:ACF}).

The potential contamination of the variability catalog by non-variable stars is primarily dominated by stars that are best characterized by a single-sinusoidal function. We split the contamination level of the stars identified by a single-sinusoidal function into two groups based on whether they have greater than or less than 0.1 normalized power. There are \nSineLow\ stars with $<$0.1 normalized power in the variability catalog, where we estimate a maximum of 27\% contamination by non-variables. The \nSineSig\ stars that are characterized by a single-sinusoidal function and have $>$0.1 normalized power are estimated to be 12\% contaminated by non-variable stars. Stars that are best characterized by either a double-sinusoidal function or an ACF are assumed to all be true variable stars. Overall, we estimate that the variability catalog as a whole is at most 21\% contaminated by non-variable stars.
%
%
%


\section{Discussion}\label{sec:discussion}

\subsection{Demographics of Variable Stars} \label{sec:stats}

The analysis presented here makes use of the \tess\ time-series photometry and the estimated stellar properties from the TICv8. The photometric analysis only considers periodic variability in the \tess\ bandpass that can be detected using the Lomb-Scargle periodogram and the auto-correlation function.  These diagnostics are not sufficient for a {\it detailed} demographic study linking the morphology of periodic variability to the specific mechanisms of stellar structure and evolution.  They do not incorporate many other ways to characterize stellar variability, such as stochastic (nonperiodic) variability, chromatic changes, spectroscopic variability, or other measures of changing flux. Nevertheless, we can extract information from this analysis that tells us about broad classes of variability by visually exploring the summary figures produced through our analysis (e.g., Figures~\ref{fig:ex_1peak}--\ref{fig:ex_ACF}) using the FilterGraph data visualization tool.\footnote{\url{https://filtergraph.com/tessvariability}} We first consider the distribution of variability across the Hertzsprung-Russell diagram\footnote{White dwarf stars were intentionally excluded from the Candidate Target List \citep[CTL;][]{Stassun18, Stassun19} during the \tess\ Prime Mission, such that they do not have 2-min cadence photometry available for the variability analyses presented in this work. A cut in luminosity and radius was also applied to the CTL in order to remove most red giant stars, with some degree of contamination expected when including all bright stars with $T_{\mathrm{mag}}<6$\,mag.} (HRD), and we then explore the relationship between periodicity and stellar luminosity.

\begin{figure*}
\epsscale{1.1}
\plotone{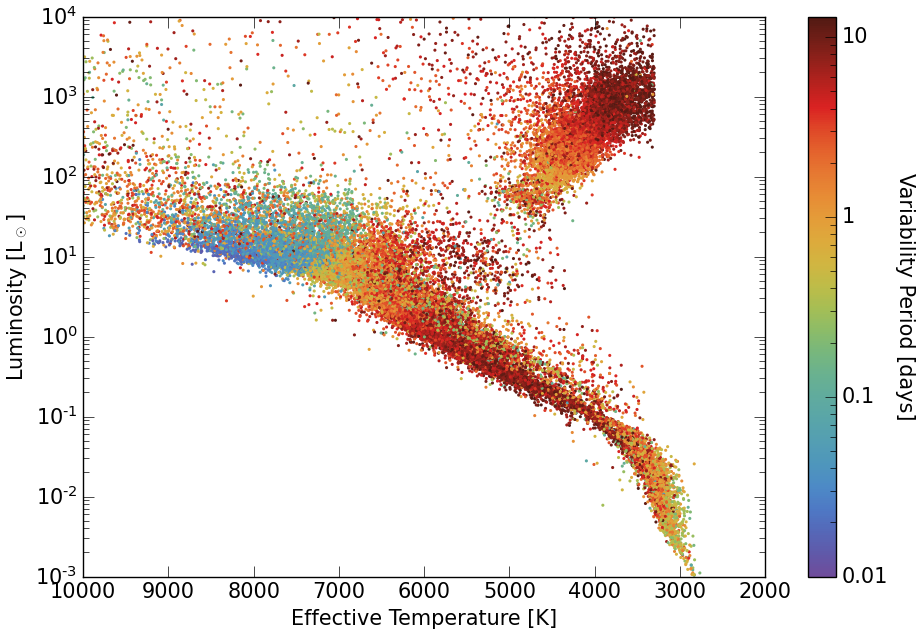}
\caption{Hertzsprung-Russell diagram of stars that are identified as significantly variable, and thus are included in the variability catalog (Tables~\ref{tab:1peak}--\ref{tab:ACF}). The points are colored by the measured variability period. Luminosities are calculated from the effective temperatures and stellar radii available in the TICv8 catalog \citep{Stassun18,Stassun19}.}
\label{fig:HRdiag4}
\end{figure*}

\autoref{fig:HRdiag4} shows the stars in the variability catalog on the HRD colored by their measured variability periods. The figure includes the subset of variable stars that were best characterized by a single-sinusoidal model (\autoref{tab:1peak}), double-sinusoidal model (\autoref{tab:2peak}), or ACF (\autoref{tab:ACF}). The location of stars on the HRD is significantly correlated with their measured variability period, which supports the expectation that the variability in these stars is primarily driven by starspot rotation and pulsations. We can describe the mechanisms for variability by connecting the different regions on the HRD with the variability periods measured from this work and examples of their observed light curves (e.g., Figures~\ref{fig:ex_1peak}--\ref{fig:ex_ACF}).

The stars with the longest variability periods are on the red giant branch and single FGK dwarf stars, which tend to have slower rotation periods as they age \citep{Irwin09, Meibom09, van_Saders13, Angus19}. Low-mass M dwarf stars ($<3500$\,K), on the other hand, have fully convective interiors such that they do not spin down as quickly over time and maintain generally short rotation periods \citep[$<$1\,day;][]{Chabrier97, Wright11, Wright18, Astudillo-Defru17, Newton17}. The shortest variability periods are found in OBAF oscillating stars, which experience multiple modes of pulsations that reveal information about their internal structures \citep{Gaia_Collaboration22, Kurtz22}. A few stars in this sample also outline the binary sequence \citep[e.g.,][]{Gaia_Collaboration18, Gaia_Collaboration19}, which is offset from the main sequence towards higher luminosities and was typically better characterized by an ACF with shorter variability periods. For stars characterized by an ACF, the main sequence tends to be dominated by eclipsing binaries, overcontact binaries, other tidally locked binaries, and fast rotating young main sequence stars, with just a few stars dominated by longer period starspot rotational modulations on the lower luminosity edge of the main sequence. The red giant branch includes a mix of intrinsic variables and eclipsing binaries. There are also a few variable stars identified along the instability strip---including those that are consistent with being RR Lyrae or Cepheid variables. 

Finally, we highlight the existence of what appears to be a prominent sub-subgiant population above the binary main-sequence and below the sub-giant branch. Sub-subgiant (SSG) stars have been recognized as likely representing unusual stellar evolution pathways ever since their initial detection as anomalies in the color-magnitude diagrams (CMDs) of some open clusters \citep[see, e.g.,][and references therein]{Mathieu03}. Subsequent studies of SSGs in clusters have proffered several possible interpretations for these systems: mass transfer in a binary system, collision of two main sequence stars, mass loss of subgiant envelopes through dynamical encounters, and reduced luminosity due to the strong surface coverage of magnetic starspots \citep[see, e.g.,][]{Leiner17}. Some recent works have concluded that mass transfer and dynamical formation pathways are disfavored based on the small numbers of SSGs in open clusters, preferring instead the strong starspot interpretation \citep[e.g.,][]{Gosnell22}. However, attempts to identify and characterize the broader SSG population in the field have only very recently begun \citep{Leiner22}. Thus, the large population of apparent SSGs in the field identified in \autoref{fig:HRdiag4} could be an opportunity to make substantial new progress in understanding these enigmatic systems.

\begin{figure*}
\epsscale{1.1}
\plotone{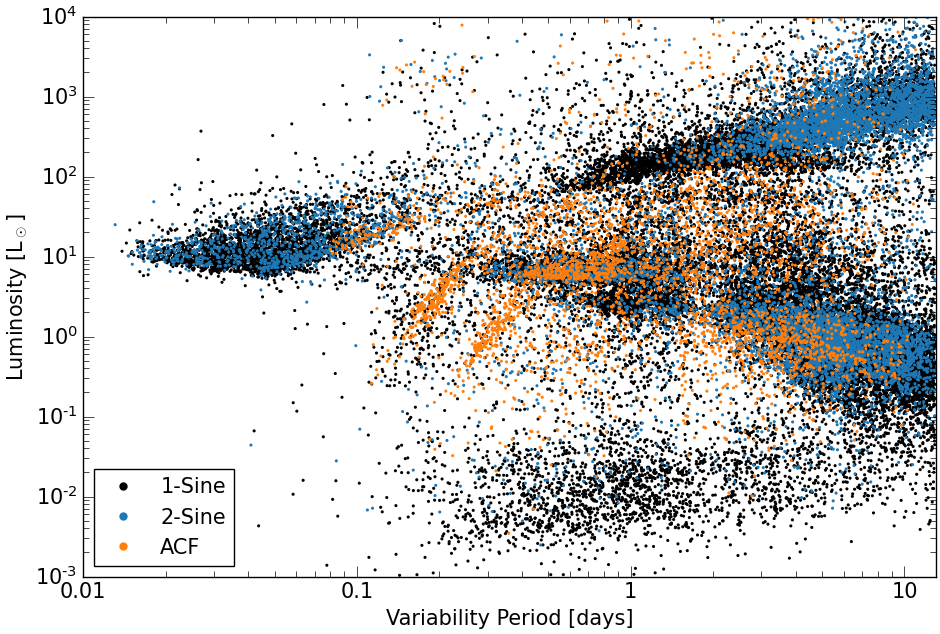}
\caption{Calculated stellar luminosities versus the measured variability periods of stars that are identified as significantly variable, and thus are included in the variability catalog (Tables~\ref{tab:1peak}--\ref{tab:ACF}). The points are colored by whether their light curves were best characterized by a single-sinusoidal function (black points), double-sinusoidal function (blue points), or ACF (orange points). Luminosities are calculated from the effective temperatures and stellar radii available in the TICv8 catalog \citep{Stassun18,Stassun19}.}
\label{fig:per-lum-type}
\end{figure*}

\begin{figure*}
\epsscale{1.1}
\plotone{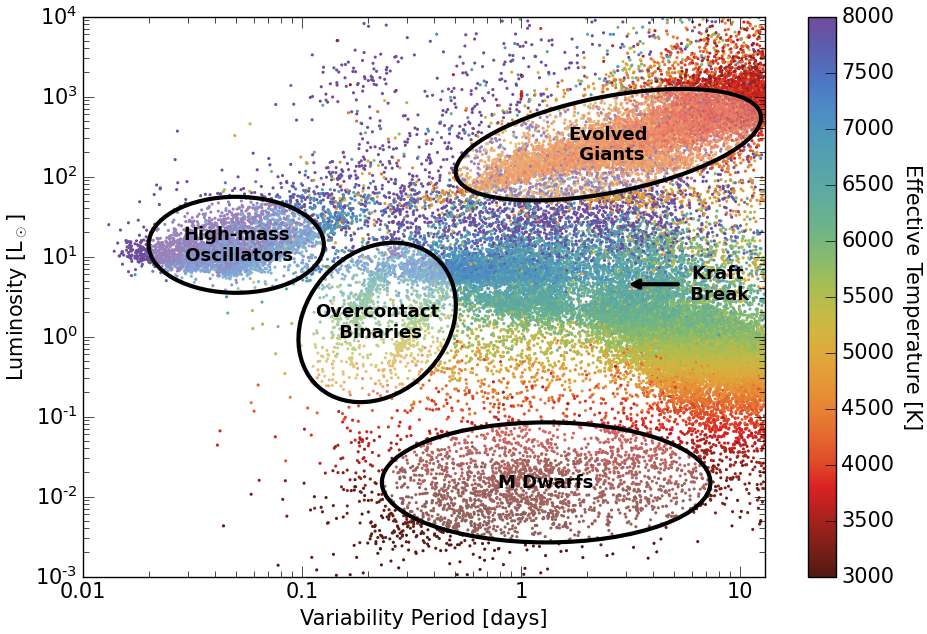}
\caption{Same as \autoref{fig:per-lum-type}, but the points are colored by their effective temperatures. Several known astrophysical relationships are highlighted and labeled with faded ellipses.}
\label{fig:per-lum-teff}
\end{figure*}

\begin{figure*}
\epsscale{1.1}
\plotone{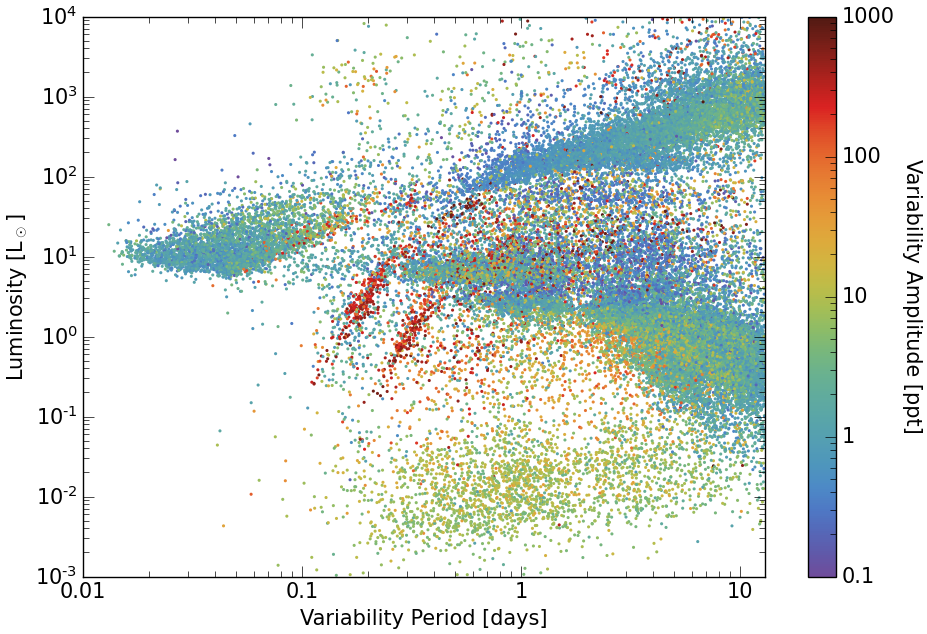}
\caption{Same as \autoref{fig:per-lum-type}, but the points are colored by the variability amplitude measured from their representative light curves.}
\label{fig:per-lum-amp}
\end{figure*}

Figures~\ref{fig:per-lum-type}--\ref{fig:per-lum-amp} show the period-luminosity relationship for the stars that are included in the stellar variability catalog. \autoref{fig:per-lum-type}, in particular, highlights which stars are best characterized by a single-sinusoidal model (black points), double-sinusoidal model (blue points), or ACF (orange points). \autoref{fig:per-lum-teff} is colored by the effective temperature and includes labels for several known astrophysical features, and \autoref{fig:per-lum-amp} is colored by the measured variability amplitude from their light curves to emphasize how the strength of the photometric variations can be used to differentiate between different types of stellar variability. 

The distribution of variable stars in the period--luminosity diagram is similar between the stars that were best characterized with a single- or double-sinusoidal function. The OBAF main sequence stars show a positive correlation between their luminosities and short-period oscillations \citep[$\lesssim$0.2\,days;][]{Gaia_Collaboration22, Kurtz22}, and we attribute the bimodal clustering of this group to ambiguous identification between the first two harmonics of oscillations that exhibit similar strength in LS normalized power. The discontinuity at $\sim$4\,\Lsun\ luminosity is attributed to the Kraft break \citep[$\sim$6200\,K;][]{Kraft67}, where cooler stars tend to rotate more slowly as magnetized stellar winds remove angular momentum from the star. There is also a positive linear relationship between luminosity and variability period for stars on the evolved giant branch ($\gtrsim$100\,\Lsun), which is consistent with asteroseismic studies of red giants \citep{Huber11, Mosser12}. The primary difference between the stars best characterized with a single-sinusoidal versus double-sinusoidal function can be seen in the giant stars, where multiple periodicities (i.e., 2-Sine) shorter than $\sim$1.5\,days are not observed in giant stars. The transition from single-sinusoidal to double-sinusoidal variability towards higher luminosity giant stars is in contrast with previous studies that have found instead that higher luminosity giants tend to pulsate with fewer distinct modes than low-luminosity giant stars \citep{Dennis14, Yu20}. However, this discrepancy may be caused by a detection bias in our catalog since lower luminosity giants tend to have smaller oscillation amplitudes. 


The period-luminosity relationship for stars characterized with an ACF (orange points in \autoref{fig:per-lum-type}) is vastly different when compared to the stars with variations that are more sinusoidal in nature (black and blue points). Stars with higher amplitude variations (redder points in \autoref{fig:per-lum-amp}) tend to be eclipsing or overcontact binaries. In particular, the period-luminosity relationship for overcontact binaries can be seen twice, where the shorter period relationship is half the binary orbital period. In order to avoid incorrectly shifting non-binary periods, we choose to not artificially double the periods for the stars that fall along the shorter period overcontact binary sequence. The linear relationship at higher luminosities ($\gtrsim$5\,\Lsun) with periods ranging between $\sim$0.01--0.2\,days represents the period-luminosity relationship for $\delta$ Scuti stars \citep{King91, Antoci19, Ziaali19, Barac22, Kurtz22}. There is also a separation in luminosity ($\sim$4\,\Lsun) for stars with lower amplitude variations that can be attributed to the Kraft break. 

\begin{figure*}
\epsscale{1.1}
\plotone{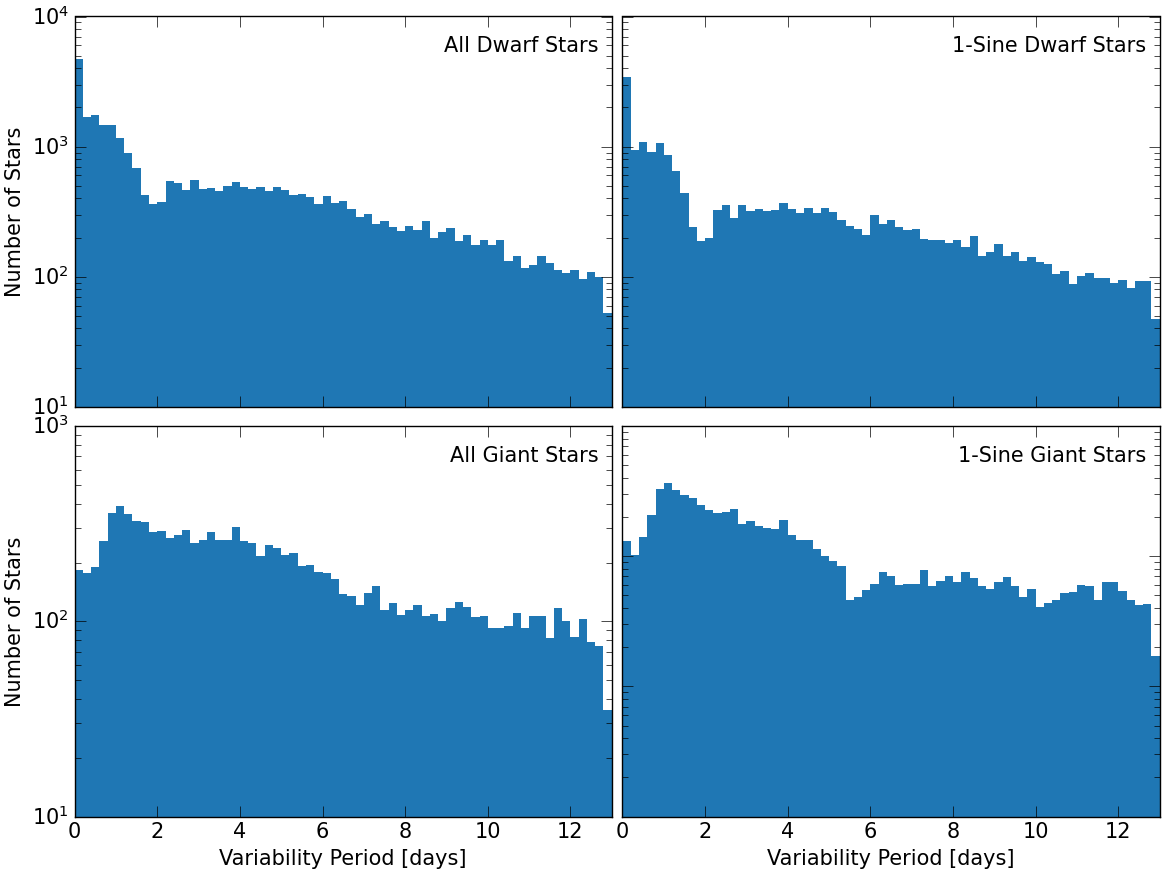}
\caption{Histograms of the measured variability periods for all (\textit{left panels}) dwarf (\textit{top panels}) and giant (\textit{bottom panels}) stars and the subset of stars best characterized by a single-sinusoidal function (\textit{right panels}). Both dwarf and giant stars characterized by a single-sinusoidal function have fewer variables at $\sim$5--6\,days due to the removal of true intrinsic variables when accounting for \tess\ systematics (see \autoref{sec:consideration}). However, only dwarf stars show fewer variable stars at $\sim$1.5--2\,days.}
\label{fig:Pvar_hist}
\end{figure*}

The left panels of \autoref{fig:Pvar_hist} shows histograms of the variability periods measured for all dwarf and giant stars in the variability catalog, where we assume the identification of dwarf and giant stars using the \texttt{LUMCLASS} flag in the TICv8 catalog \citep{Stassun18, Stassun19}. There are significantly fewer dwarf stars that exhibit variations on timescales of 1.5--2\,days, which is not necessarily related to the Kraft break since these rotation periods are also uncommon for M dwarf stars (see \autoref{fig:per-lum-teff}). If the dearth of variability at 1.5--2\,days were caused by a systematic effect, then the drop in occurrence would be observable in both the dwarf and giant stars. On the contrary, the drop in variable stars at 1.5--2\,days only occurs in dwarf stars even when only the subset of stars that were best fit by a single-sinusoidal function are considered (right panels; \autoref{sec:consideration}). \citet{Kounkel22} suggested that binaries amount FGK-type stars with periods faster than $\sim$2\,days, and M-type stars with periods faster than $\sim$1.5\,days are predominately binaries that are possibly undergoing tidal interactions. Furthermore, the lack of $\sim$2 day periodicities spans over 4 orders of magnitude in luminosity, including M dwarfs, which suggest that it is could be related to binarity rather than pulsations. However, we leave an in-depth investigation in the binary occurrence rate in FGKM stars near variability periods of 1.5--2\,days to a future work.

In the right panels of \autoref{fig:Pvar_hist} we observe fewer variable stars at 5--6\,days for both dwarfs and giants, which is in contrast to \citet{Holcomb22} where they identified an overdensity of rotational variables at $\sim$5\,days. However, they suspected that their observed overdensity of variable stars at $\sim$5\,days could be related to uncorrected systematics in the \tess\ PDCSAP photometry. This interpretation is consistent with how we removed variable stars from the variability catalog that were best be characterized by a single-sinusoidal function but could be attributed to spacecraft systematics---including some stars that are truly variable at 5--6\,days (see \autoref{sec:consideration}).

\subsection{Demographic Comparisons to Similar Work}

Projects similar to this paper have been undertaken using photometric data from other sky surveys, as described in \autoref{sec:intro}.  The observational and methodological heterogeneity of these surveys makes it hard to directly compare their bulk variability results.  However, we can qualitatively compare the results from this study to those of the two of the most similar surveys: the exploration of variable stars across the HRD using \textit{Gaia} data in \citet{Gaia_Collaboration19}, and the search for periodic stellar variability in data from the ground-based NGTS survey in \citet{Briegal22}.  The observational properties of both surveys have some overlap with the stellar population observed by \tess, and we intend to do direct star-by-star comparisons with those results in future work (see discussion in \autoref{sec:future}).

In \citet{Gaia_Collaboration19}, the authors examine the position of variable star classes on the HRD, separated by the various intrinsic and extrinsic types of known stellar variability.  They compile a large set of known variable stars from various published sources and locate them on the HRD using \textit{Gaia} colors and magnitudes, and also conduct a limited exploration of stellar variables using \textit{Gaia} photometry, considering only the amplitude (the interquartile range) of the \textit{Gaia} $G$-band light curves.  They identify regions of the HRD with a greater incidence of variability (see their Fig. 8), and also differences in typical variability amplitude across the HRD (their Fig. 9). \citet{Briegal22} conduct a general search for periodic variability in their lightcurves, which cover a set of fields around the sky, each observed for an average of 141 nights across an average time baseline of 218 days.  Their variability search method shares some broad similarities to ours, while also being careful to remove the systematic trends that arise in wide-field ground based photometric observing.  They also plot their variability detections across a CMD using \textit{Gaia} magnitudes. 

We can examine some of the variability distributions across the HRD between the studies.  This comparison is only qualitative and approximate, since the \textit{Gaia} and NGTS analyses use CMDs whereas we are plotting physical properties in an HRD.  Also, the selection of targets for \tess\ 2-min cadence observations was motivated by the search for transiting planets and not a broad sample of stars in a magnitude-limited way, which will certainly lead to selection biases in the numerical results. \textit{Gaia} observes deeper than \tess\ to fainter stars, and over a longer time frame (22 months), with fewer individual epochs of observation (of order 20 or more observations).\footnote{Partly because of that, we do not see the white dwarf sequence in the \tess\ data (along with the fact that the TICv8 stellar parameters were not designed to include white dwarfs).}  NGTS observes a more similar set of stars in apparent magnitude, and over a time baseline more similar to \textit{Gaia}, but with a number of epochs more similar to \tess.  On the other hand, \tess\ light curves have much greater photometric precision and duty cycles than either of the other surveys.  Nevertheless, we can still qualitatively compare our overall distributions to those in these other two papers.

\begin{figure*}
\epsscale{1.1}
\plottwo{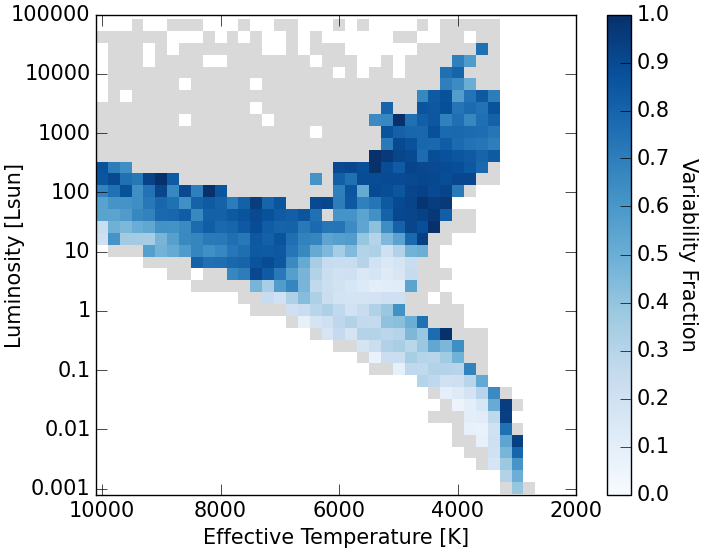}{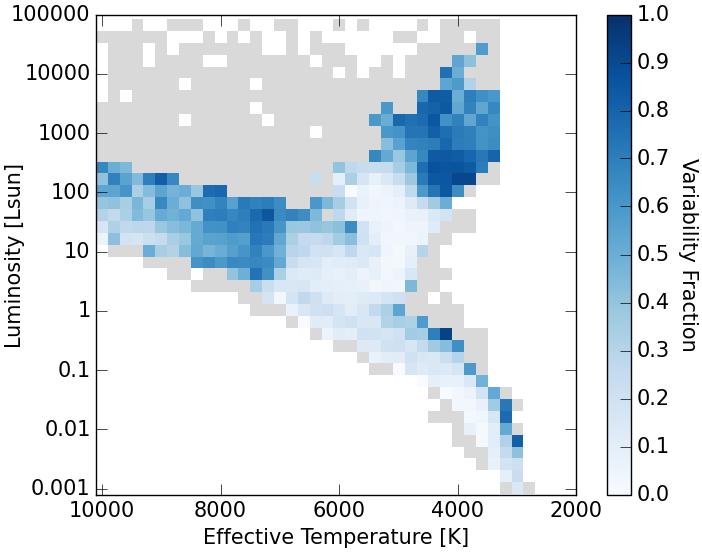}
\caption{The fraction of stars that are determined to be significantly variable across the HRD, with normalized power cutoffs of 0.01 (\textit{left}) and 0.1 (\textit{right}). The gray bins indicate areas of low completeness ($<$10 stars). Note the increased fractions of variable stars at the upper edge of the binary main-sequence as well as among the sub-subgiant population.}
\label{fig:varFrac}
\end{figure*}

We plot the fraction of stars that we identify as variable across the HRD in \autoref{fig:varFrac}.  Similar to Fig. 8 from \citet{Gaia_Collaboration19} and Figs. 3a and 7 in \citet{Briegal22}, we show higher rates of variability---especially with short periods---for the hot stars of the upper main sequence, and also along the top edge of the lower main sequence, which includes the binary sequence. The increased variability fraction for hotter stars on the upper main sequence is consistent with observations of the $\delta$ Scuti instability strip \citep{Murphy19}. We do not show the high rates of variability at the upper part of the red giant branch due to the limited time span of the single-sector \tess\ observations in this analysis.  We do show high rates of variability on the lower part of the giant branch, but the comparison of an HRD with a CMD, and varying timescales of variability probed between the different surveys make that point difficult to verify.

\begin{figure*}
\epsscale{1.1}
\plotone{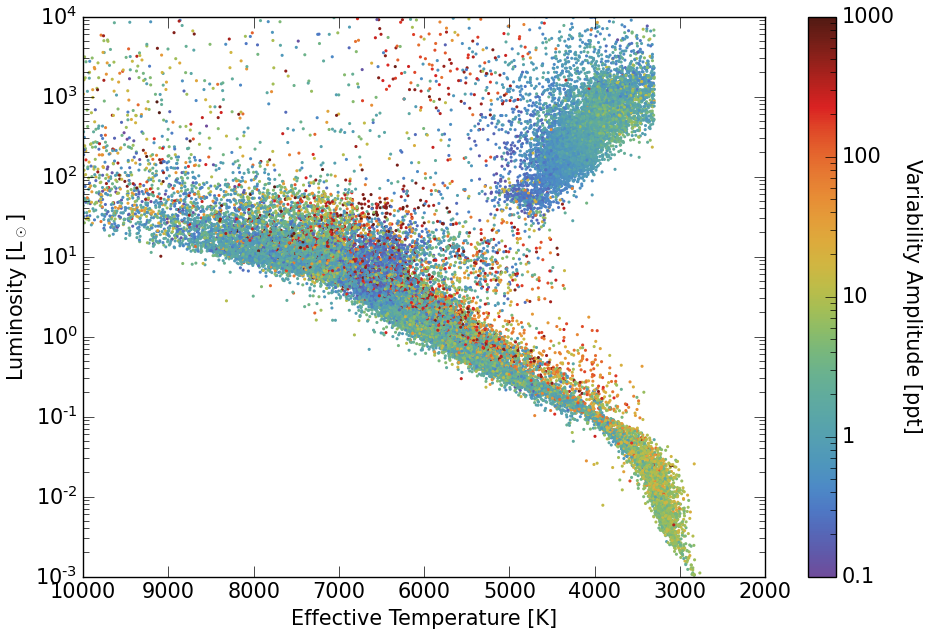}
\caption{Same as \autoref{fig:HRdiag4}, but the points are colored by the measured variability amplitude from the light curves. Note the gradient of increasing variability amplitude with increased displacement above the main-sequence; these are evidently binary stars on the binary main-sequence. Interestingly, the sub-subgiant population exhibits a heterogeneity of variabilty amplitudes.}
\label{fig:HRdiag5}
\end{figure*}

We can look at the distribution of variability amplitude between \autoref{fig:HRdiag5} and Fig. 9 in \citet{Gaia_Collaboration19}.   We show the same increase in typical variability amplitude for stars on the upper edge of the main sequence, around the binary sequence. We see a gap in variability between the main sequence and the upper end of the giant branch, although the correspondence between the gaps in the two plots around the subgiant and lower giant branch regimes is not well defined.  Above the part of the main sequence where the subgiant branch emerges, we find a number of very high-amplitude variables, which turn out to be classical pulsators in the instability strip.  We do not see the high amplitudes among the red giant stars in the \tess\ data due to the limited time frame of the current analysis, although once we incorporate multi-sector \tess\ data in future work the red giant high-amplitude, long-period variability should become more apparent. 

\begin{figure*}
\epsscale{1.1}
\plotone{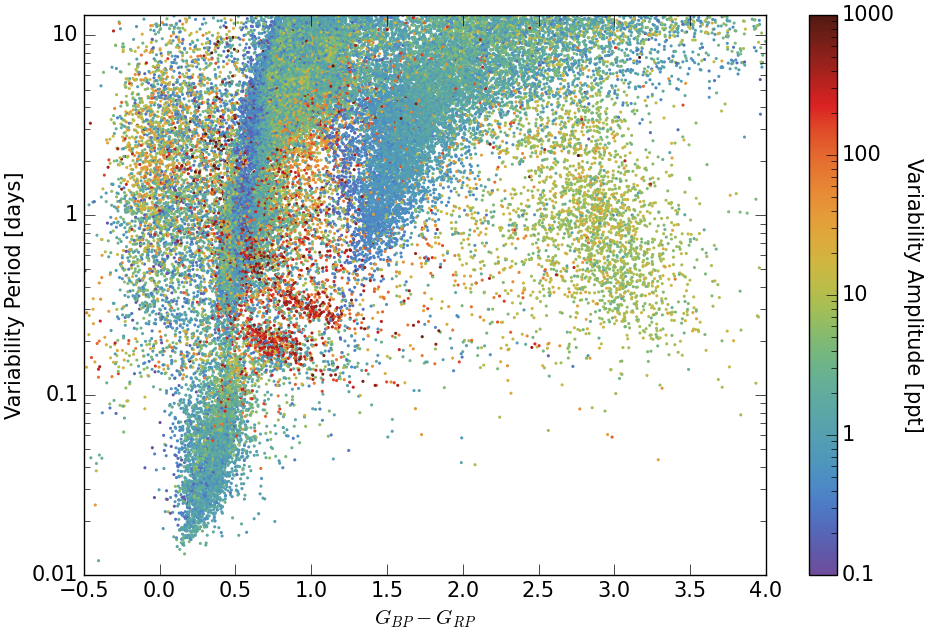}
\caption{Variability period versus \textit{Gaia} $G_{BP}-G_{RP}$ color, with the points colored by the measured variability amplitude from the light curves.}
\label{fig:NGTScomp}
\end{figure*}

While \citet{Gaia_Collaboration19} did not include an investigation of periodic variability, \citet{Briegal22} does compare the periodicity of their variables to the colors of the stars in a manner similar to our analysis throughout \autoref{sec:stats}. We make a direct comparison to the period-color distribution of Fig. 8 from \citet{Briegal22} in \autoref{fig:NGTScomp}, where we show our measured variability periods versus \textit{Gaia} $G_{BP}$ and $G_{RP}$, and color the points by the amplitude of their variability. We see a similar distribution to the figure in \citet{Briegal22}, with a gradual trend towards longer variability periods for redder stars. Also seen at periodicities between 0.1 and 0.5 days are two populations of variables with the opposite trends with color, which we visually confirm as representing the overcontact binaries.  In general, the upper population represents those stars identified at the correct orbital period, while the lower population represents those identified at half the true period, as noted also in \citet{Briegal22}.

\subsection{Comparing Variability Catalogs} \label{sec:crossmatch}
There have been a number of lists of photometric variables published by other projects (e.g. the superWASP and ASAS-SN catalogs), but the heterogeneity of photometric surveys has generally made cross-survey combination of variable lists difficult to conduct and incomplete. That tends to be due to differing observing cadences, time baselines, duty cycles, passbands, magnitude ranges, photometric precision, angular resolution, and other observational parameters. Nevertheless, the sky coverage, photometric precision, and duty cycle of \tess\ makes this data set a more natural base catalog of (periodic) variability from which to build a true all-sky variability catalog, at least for short-period variability of bright stars.

\begin{figure*}
\epsscale{1.1}
\plottwo{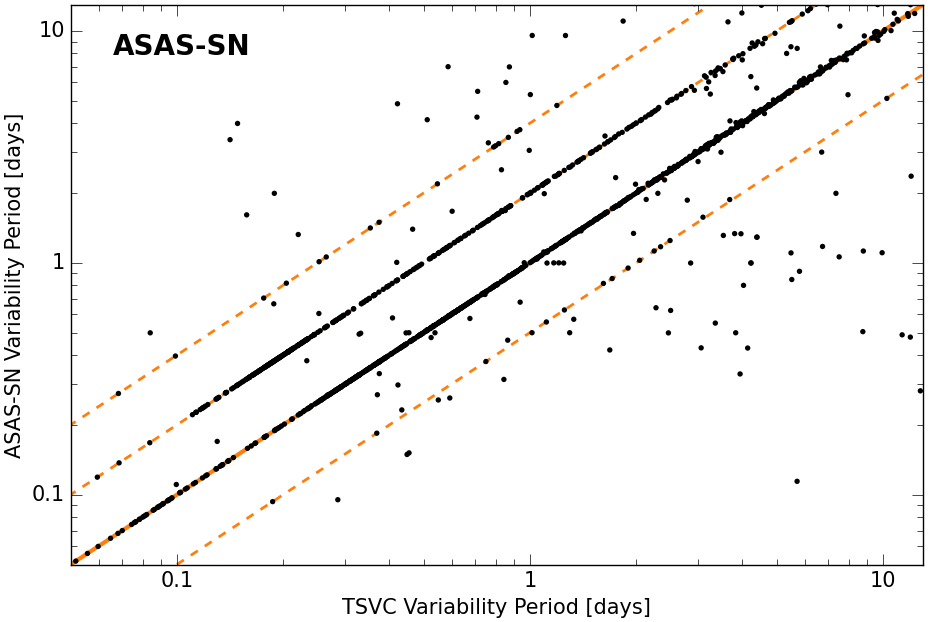}{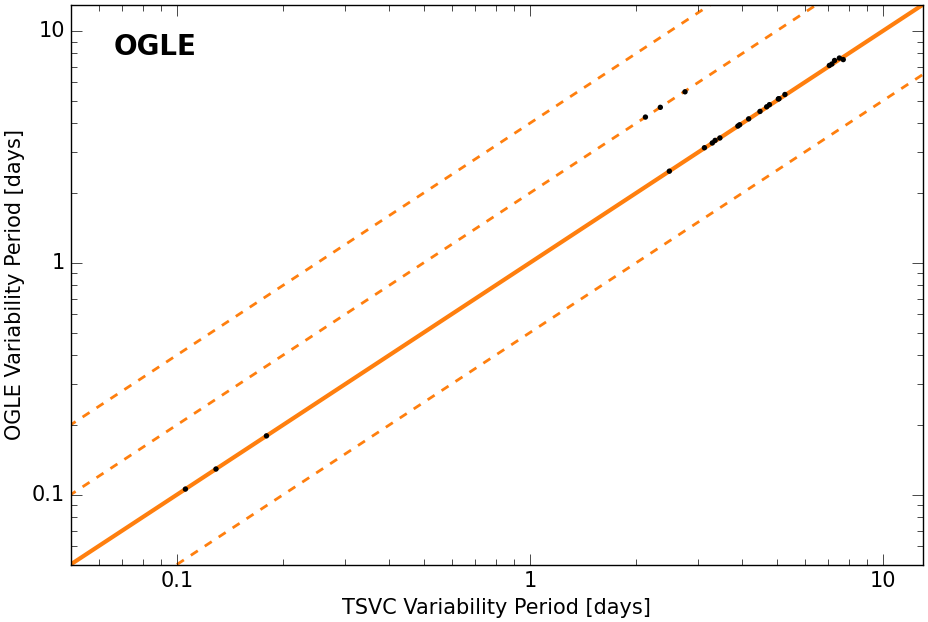}
\plottwo{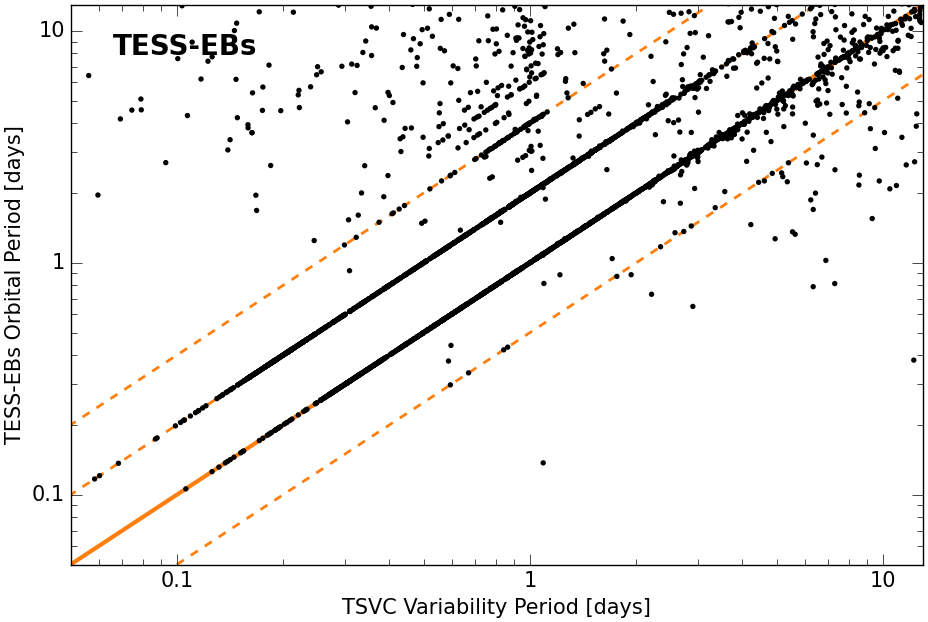}{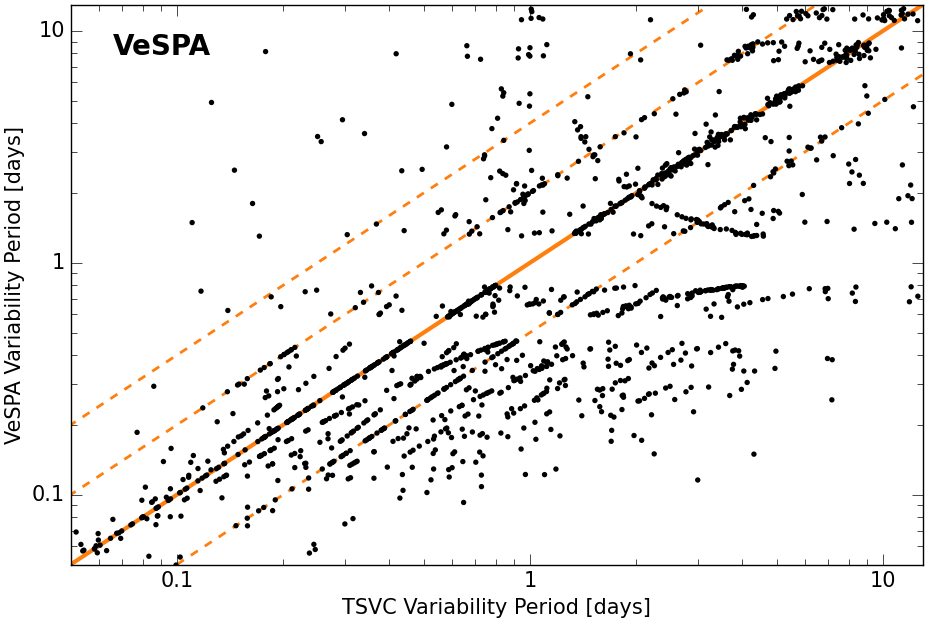}
\plottwo{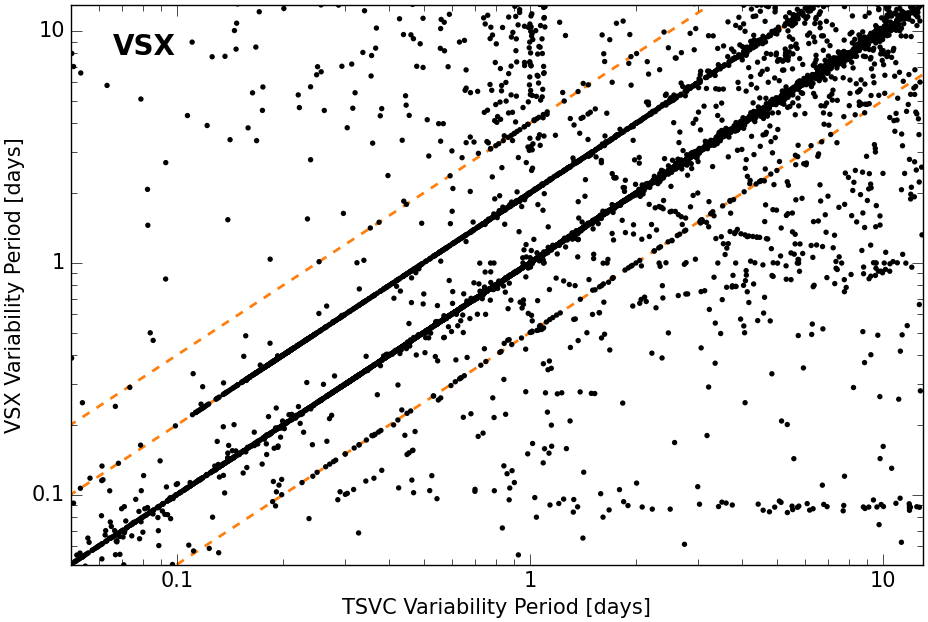}{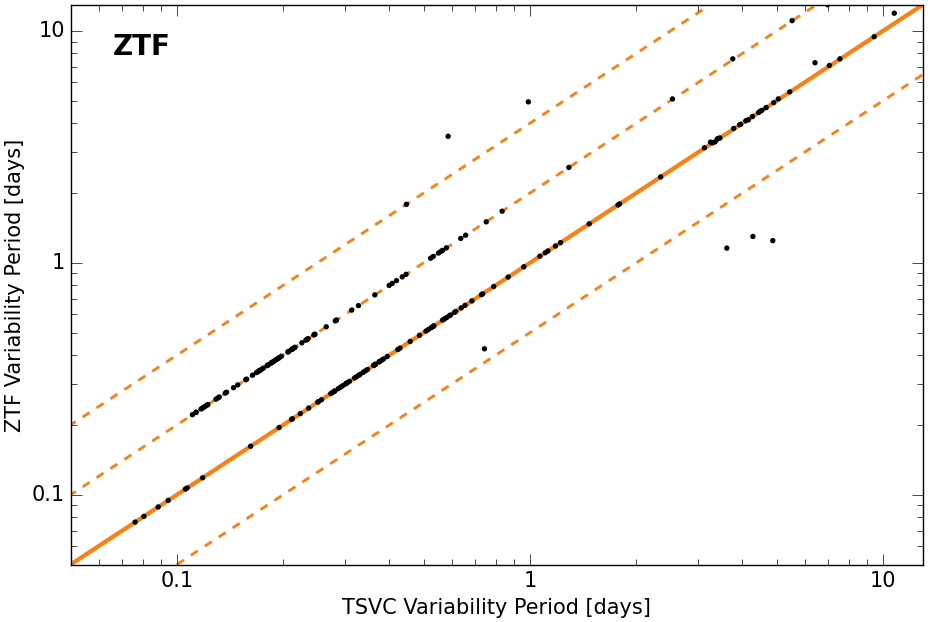}
\caption{Comparison of the measured variability periods between our variability catalog and six other variable star surveys: ASAS-SN (\textit{top left}), OGLE (\textit{top right}), TESS-EBs (\textit{center left}), VeSPA (\textit{center right}), VSX (\textit{bottom left}), and ZTF (\textit{bottom right}). A perfect match between the variability periods is indicated by the solid orange line. The 2:1 and 4:1 alias periods are indicated by the dashed orange lines.}
\label{fig:period_compare}
\end{figure*}

Here we compare the contents of variable stars in our catalog with those identified in several other ground- and space-based surveys. In particular, we select the following variability surveys and catalogs: SuperWASP: VeSPA \citep{Pollacco06, McMaster21, Thiemann21}, OGLE \citep{Udalski08}, ASAS-SN \citep{Jayasinghe18}, \textit{Gaia} DR3 \citep{Gaia_Collaboration21, Eyer22}, ZTF \citep{Yao19, Chen20}, TESS-EBs (\citealt{Prsa22}; \dataset[DOI: 10.17909/t9-9gm4-fx30]{http://dx.doi.org/10.17909/t9-9gm4-fx30}), and the AAVSO International Variable Star Index\footnote{\url{https://www.aavso.org/vsx/}} (VSX). In order to evaluate our efforts at identifying variable stars, we only compare the stars from these surveys that overlap with our base sample of stars where we searched for periodicity in their 2-min light curves obtained during \tess\ Sectors 1--26. For the \textit{Gaia} DR3, ASAS-SN, and TESS-EB catalogs we are able to match targets using their TIC or \textit{Gaia} identifiers. Otherwise, we perform a 1\arcsec\ search radius based on their RA and Dec. \autoref{tab:crossmatch} shows the total number of stars that were successfully cross-identified from each survey compared to our base sample ($N_\mathrm{tot}$) and the number of stars that were also detected as variable in our catalog ($N_\mathrm{var}$). In total, there are 18,734 unique stars that overlap between this catalog and these existing catalogs, which implies that the catalog that we present in this work includes $\sim$65,000 new variable stars. We show how the measured variability periods compare between our variability catalog and each of the other surveys in \autoref{fig:period_compare} (excluding \textit{Gaia} DR3).\footnote{The \textit{Gaia} DR3 variability periods are paired with their classifications. We aim to simply evaluate our period measurements compared to those obtained through other surveys, which we are able to sufficiently assess using the other variability catalogs.} In addition to highlighting matching variability periods (solid orange line), \autoref{fig:period_compare} also highlights common harmonic aliases where our variability periods are different from those reported in other catalogs by factors of two or four (dashed orange lines). Harmonic aliases are especially often found for eclipsing binaries in cases where the shape and depth of the secondary eclipse closely matches that of the primary eclipse. Our variability search often identifies eclipsing binaries at half of their actual orbital period when the shape and depth of the secondary eclipse closely matches that of the primary eclipse. When comparing variability periods derived in this paper with eclipsing binary orbital periods reported by the TESS-EBs catalog (center left panel), we find that our analysis tends to select shorter-period harmonic aliases that can be many times shorter than the orbital period. Of note are some instances of eccentric eclipsing binaries that are identified as being variable at odd harmonic aliases (e.g., three or five times shorter). The identification of eclipsing binaries at certain eccentricities is a by-product of the LS periodogram assigning high LS power to the unequally spaced primary and secondary eclipses. The variability period comparisons with variability catalogs derived from ground-based observational surveys (i.e., VeSPA and VSX) in \autoref{fig:period_compare} reveal arc-like features, which are caused by aliases between the true signal and the diurnal observing window \citep[see Fig. 24 in][]{VanderPlas18}. Besides these well-understood aliases, there is overall good agreement between the variability periods in our catalog and those reported by other stellar variability surveys. For the variability periods measured to be $<$13\,days, 93\%, 100\%, 84\%, 54\%, 60\%, and 97\% of the variability periods reported in the ASAS-SN, OGLE, TESS-EBs, VeSPA, VSX, and ZTF catalogs, respectively, agree within 10\% of our variability period measurement, or are within 10\% of twice or half of our variability period measurement. However, the agreement between our variability period measurements and those reported in the ASAS-SN, OGLE, and ZTF catalogs is much better, with a median scatter that is less than 1\%.

\begin{deluxetable}{lcc}
\tablecaption{Number of Targets Matched Between Stellar Variability Catalogs\label{tab:crossmatch}}
\tabletypesize{\scriptsize}
\tablehead{
\colhead{Survey Name} & \colhead{$N_\mathrm{tot}$\tablenotemark{a}} & \colhead{$N_\mathrm{var}$\tablenotemark{b}}
} 
\startdata
ASAS-SN & 9830 & 8278 \\
\textit{Gaia} DR3 & 14309 & 10281 \\ 
OGLE & 35 & 33 \\
TESS-EBs & 3692 & 3365 \\
VeSPA & 2132 & 1872 \\
VSX & 13398 & 10741 \\ 
ZTF & 286 & 256
\enddata
\tablenotetext{a}{Overlapping targets that were analyzed in our variability search.}
\tablenotetext{b}{Targets that are also identified as variable in the catalog presented in this work.}
\end{deluxetable}

\begin{figure*}
\epsscale{1.1}
\plottwo{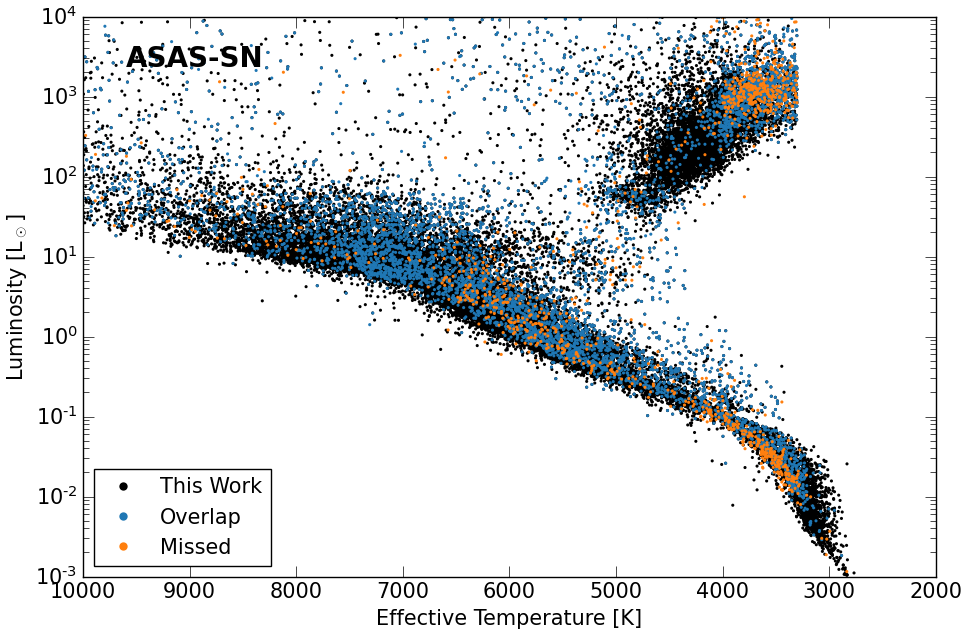}{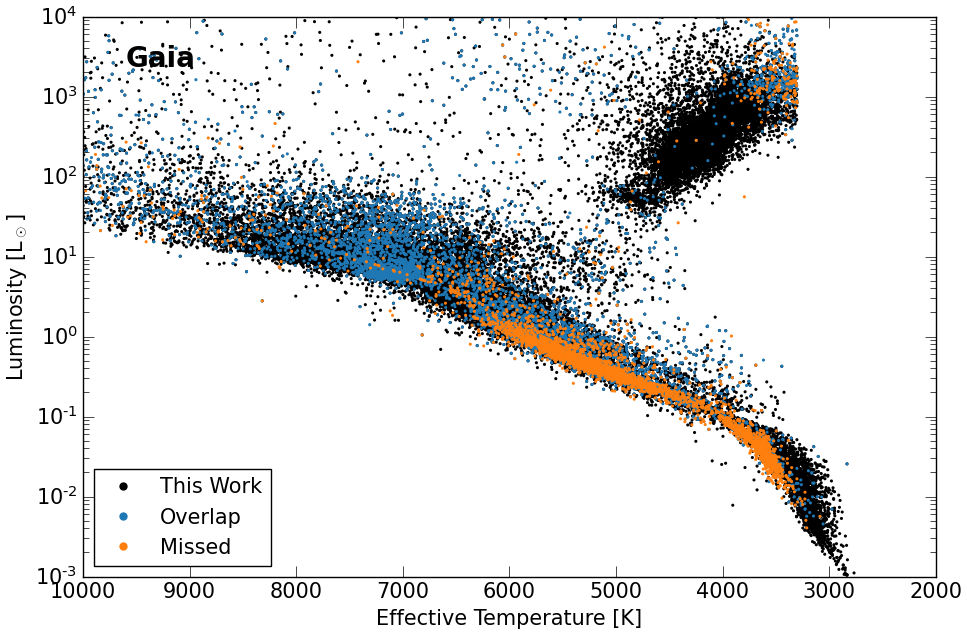}
\plottwo{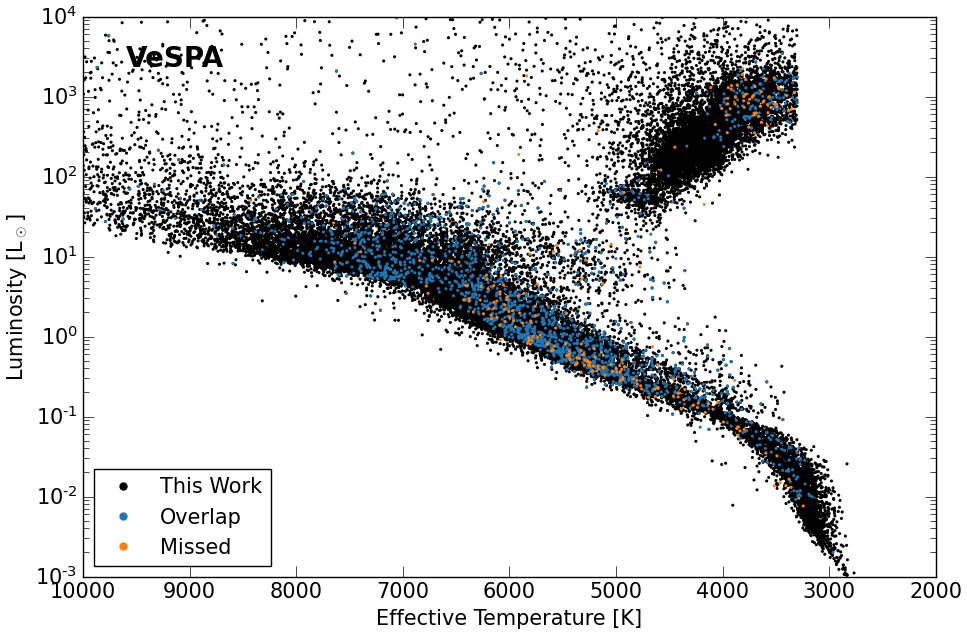}{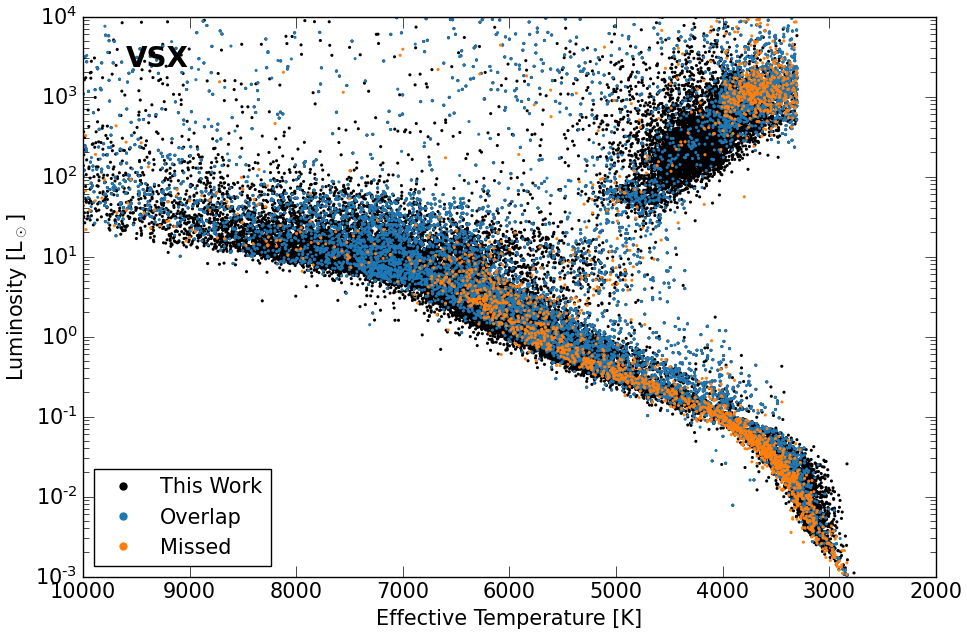}
\caption{The locations of stars on the HR diagram for those matched between our analyzed sample and four other variable star surveys: ASAS-SN (\textit{top left}), \textit{Gaia} DR3 (\textit{top right}), VeSPA (\textit{bottom left}), and VSX (\textit{bottom right}). All stars included in our variability catalog are shown by the black points. The stars that are variable in the respective survey and are also included in our variability catalog are indicated by the blue points. The stars that are identified as variable in the compared survey, but \textit{not} identified in this work are highlighted by the orange points.}
\label{fig:HRdiag_compare}
\end{figure*}

\begin{figure*}
\epsscale{1.1}
\plotone{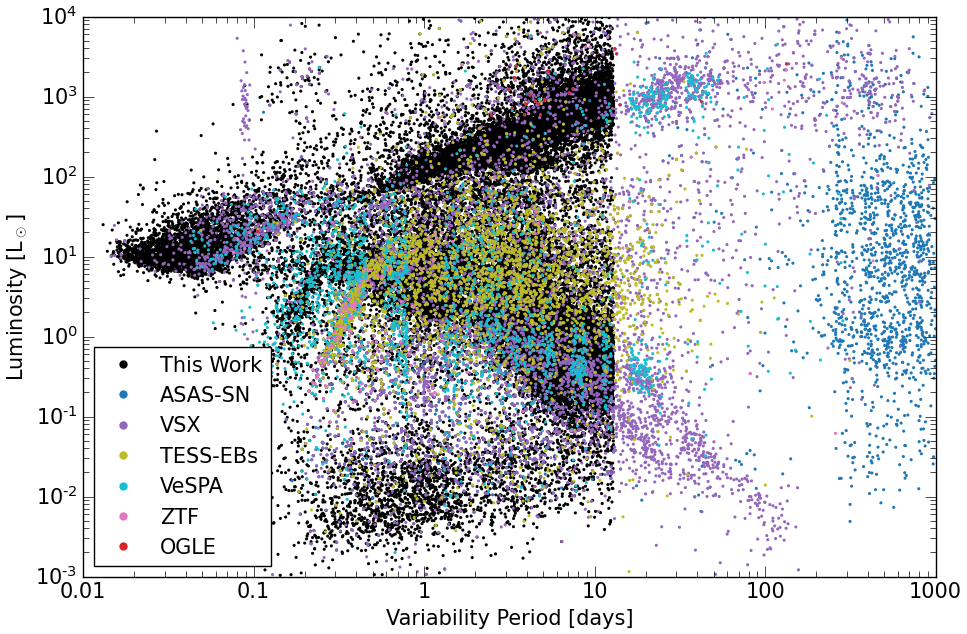}
\caption{Same as \autoref{fig:per-lum-type}, but the colored points indicate that they are found in other variability catalogs.}
\label{fig:per-lum_compare}
\end{figure*}

In \autoref{fig:HRdiag_compare}, we show how variable stars from ASAS-SN, \textit{Gaia} DR3, VeSPA, and VSX are distributed across the HRD. The stars that are identified as significantly variable in both our catalog and the respective catalog are indicated by the blue points ($N_\mathrm{var}$ in \autoref{tab:crossmatch}). The stars listed as variables in the other catalog that are missed by our catalog are indicated by the orange points. \autoref{fig:per-lum_compare} shows the period-luminosity relationship for all variable stars identified in the selected catalogs (excluding \textit{Gaia} DR3). It can be seen that many stars exhibit variability at periods longer than our 13-day search limit, primarily due to the longer observing baselines of ground-based observations available to ASAS-SN, VSX, and VeSPA. ASAS-SN especially excels at identifying variable stars with periodicities on timescales of several hundred days, and the VSX and VeSPA catalogs fill in the parameter space for periodic variability at 10--100\,days. Most notably, the variable stars from the VSX and VeSPA catalogs further extend the linear trends observed between luminosity and variability period ($\sim$0.05--100\,days), where higher luminosity giant stars and lower luminosity main sequence stars both tend to have longer variability periods (presumably their rotation periods) compared to lower luminosity giants and higher luminosity (i.e., hotter) main sequence stars. Stars with longer variability periods ($>$13\,days) could largely account for the stars that did not make the cut into our variability catalog despite being included in our LS periodogram search (orange points in \autoref{fig:HRdiag_compare}). The TESS-EBs catalog largely identifies stars along the tight period-luminosity relationship for over-contact binary systems, but also identifies many detached binaries with orbital periods of 0.5--20\,days that span luminosities of 1--100\,\Lsun. The luminosities, in the case of binary systems, are still calculated using the reported effective temperature and stellar radius in the TICv8 catalog and assuming a single star, which results in binaries being offset from the main sequence towards higher luminosities in the HRD \citep[e.g.,][]{Gaia_Collaboration18, Gaia_Collaboration19}. The ZTF catalog primarily identifies over-contact binary systems, with a few rotational variables. There are very few stars that overlap with our sample from the OGLE survey, but the majority are high-luminosity giant stars. The \textit{Gaia} DR3 variability periods are not included in \autoref{fig:per-lum_compare}. However, based on the HRDs shown in \autoref{fig:HRdiag_compare}, it appears that \textit{Gaia} DR3 variable stars cover a similar parameter space as the VSX catalog.

Compared to these other variability catalogs, our methodology and use of \tess\ photometry tends to identify new variable stars at the high- and low-temperature ends of the main sequence, and lower luminosity giant stars. The 2-min cadence continuous-stare strategy of \tess\ is especially ideal for identifying photometric variability on short timescales and at a higher precision than is possible for ground-based surveys, such that new variable stars with short periodicities ($\ll$1\,day) or that are relatively faint (e.g., M-dwarfs) should not come as a surprise even when compared to extensive community-driven efforts such as VSX.

We recognize that this is not an exhaustive list of all surveys that observed these stars---especially given that \tess\ prioritized bright, nearby dwarf stars across nearly the entire sky suitable for the detection of small transiting exoplanets. However, we have demonstrated that there is sufficient overlap and consistent variability period measurements between our catalog and several other variability catalogs from the literature. Furthermore, our catalog includes many new variables that were likely beneath the photometric amplitude detection limit of ground-based surveys or periodic with a cadence that is aliased with ground-based observing strategies.

\subsection{Future Directions} \label{sec:future}
The work presented in this paper is only an initial step in detecting and analyzing stellar variability using \tess\ data.  There are numerous ways to expand upon this work.  A natural next step that we are already working on is to extend the periodicity search beyond 13 days by stitching \tess\ observations across multiple sectors from the Simple Aperture Photometry, which will be especially powerful for examining stars in the \tess\ CVZs.  Furthermore, there are now multiple pipelines extracting light curves of stars not selected for 2-min cadence observation, but are observed in the FFIs.  These include light curves from QLP \citep{Huang20}, TESS-SPOC \citep{Caldwell20}, DIA \citep{Oelkers18}, and DIAMANTE \citep{Montalto20}, along with customized extraction tools for individual targets such as \texttt{lightkurve} \citep{Lightkurve_Collaboration18}, \texttt{eleanor} \citep{Feinstein19}, and \texttt{giants} 
\citep{Saunders22}.  The availability of truly complete magnitude-limited stellar samples available from the FFIs would allow a future population analysis of variability unaffected by the target selection process for the 2-min cadence targets.\footnote{Although even the FFI data will be constrained by magnitude limits of the \tess\ aperture, and blending due to the \tess\ pixel size.} For example, \citet{Kounkel22} recently explored rotational variability that is detectable from the FFIs, which would be a valuable work for comparison when exploring the FFIs for broad variability.

An additional extension of this work is to combine the \tess\ light curves with photometry from other wide-field surveys.  Ground-based surveys such as ASAS \citep{Pojmanski02}, HATNet \citep{Hartman04}, KELT \citep{Oelkers18}, SuperWASP \citep{Pollacco06}, ASAS-SN \citep{Jayasinghe18}, and ZTF \citep{Yao19} cover large fractions of the sky with broadband photometry (typically in a single bandpass) over long periods of time.  While the data from these surveys do not reach the photometric precision or duty cycle of \tess, they do cover much longer time frames.  Combining \tess\ photometry with ground-based archival data will allow us to search for longer-term variations or changes in variability. 

Lastly, there are a number of efforts underway to classify photometric variables with much finer physical and observational categories than has been done in this work.  Notable examples are the OGLE project \citep{Udalski08}, the \textit{Gaia} DR3 variability search \citep{Gaia_Collaboration21, Rimoldini22}, specific types of variable stars observed with \tess\ \citep[e.g.,][]{Antoci19, Howard20a, Hon21, Avallone22, Barac22, Holcomb22, Prsa22, Kurtz22, Saunders22}, and many others.  The \tess\ Asteroseismic Science Operations Center (TASOC) is engaged in much more detailed and individualized analysis of certain types of stellar variability \citep{Audenaert21}.  Additionally, both supervised and unsupervised machine learning classification tools have been deployed to automatically classify variable stars \citep[e.g.,][]{Blomme10, Audenaert21, Barbara22, Claytor22}.  We hope that the more limited variability characteristics and broad quantitative parameters we have presented here, along with the open description of our selection and threshold procedures, can be useful for such projects going forward.
%
%
%


\section{Summary} \label{sec:summary}

We used \tess\ Prime Mission 2-min cadence photometry to search for stellar variability in $\sim$200,000\,stars. Using a LS periodogram and ACF, we searched for photometric periodic variability on timescales up to 13\,days in the light curves observed from each \tess\ sector. In Tables~\ref{tab:1peak}--\ref{tab:ACF}, we present our variability catalog, which includes $\sim$46,000 stars that are considered astrophysically variable to high confidence. In \autoref{sec:stats}, we identify several trends of variability on the HRD and period--luminosity diagram. We find that the photometric variability period and amplitude of variability can be used as significant metrics for classifying different types of variability across the HRD. Furthermore, the distribution of stars in period--luminosity space reveals important information about different astrophysical mechanisms driving stellar variability. After careful consideration of variations caused by \tess\ spacecraft systematics, we also identify a dearth of dwarf star variability around 1.5--2\,days (see \autoref{fig:Pvar_hist}) where the astrophysical mechanism is not currently well-understood. 

Overall, this work presents a broad overview of stellar variability observed across nearly the entire sky, but presents many opportunities for future investigations into stellar astrophysics and how stellar variability may affect exoplanetary systems. Stellar variability has already been shown to be a source for false positive exoplanet detections \citep[e.g.,][]{Henry02, Robertson14, Robertson14-1, Robertson15, Kane16, Hojjatpanah20, Prajwal22, Simpson22}, but avoiding active stars in exoplanet studies has also led to observational biases against planets around stars of certain spectral types and stages of evolution that tend to be particularly active (Simpson et al. 2023, submitted). Since all stars are variable at some point in their lifetime, understanding how stellar activity affects the measurements of exoplanet properties is a critical step towards understanding how star-planet systems---and especially planetary atmospheres---evolve over time. 
%
%
%


\section*{}

The authors thank Guillermo Torres for helpful conversations and insight into the interpretations of this work. The authors acknowledge support from NASA grants 80NSSC18K0544 and 80NSSC18K0445, funded through the Exoplanet Research Program (XRP), and NASA grant 80NSSC20K0447 funded through the ADAP program. T.F. acknowledges support from the University of California President's Postdoctoral Fellowship Program. This work was supported by a research grant (00028173) from VILLUM FONDEN. Funding for the Stellar Astrophysics Centre is provided by The Danish National Research Foundation (Grant agreement no.: DNRF106). D.H. acknowledges support from the Alfred P. Sloan Foundation and the National Aeronautics and Space Administration (80NSSC21K0652). We acknowledge the use of public \tess\ data from pipelines at the \tess\ Science Office and at the \tess\ Science Processing Operations Center. Resources supporting this work were provided by the NASA High-End Computing (HEC) Program through the NASA Advanced Supercomputing (NAS) Division at Ames Research Center for the production of the SPOC data products. All of the data presented in this paper were obtained from the Mikulski Archive for Space Telescopes (MAST). The specific observations analyzed can be accessed via \dataset[DOI: 10.17909/t9-nmc8-f686]{http://dx.doi.org/10.17909/t9-nmc8-f686}. STScI is operated by the Association of Universities for Research in Astronomy, Inc., under NASA contract NAS5-26555. Support for MAST for non-HST data is provided by the NASA Office of Space Science via grant NNX13AC07G and by other grants and contracts. This paper includes data collected with the \tess\ mission, obtained from the MAST data archive at the Space Telescope Science Institute (STScI). Funding for the \tess\ mission is provided by the NASA Explorer Program. STScI is operated by the Association of Universities for Research in Astronomy, Inc., under NASA contract NAS 5–26555. This research made use of Lightkurve, a Python package for \kepler\ and \tess\ data analysis \citep{Lightkurve_Collaboration18}. We acknowledge with thanks the variable star observations from the \textit{AAVSO International Database} contributed by observers worldwide and used in this research.



\facilities{TESS, AAVSO}

\software{Astropy \citep{Astropy_Collaboration13, Astropy_Collaboration18},
          Astroquery \citep{Ginsburg19},
          Lightkurve \citep{Lightkurve_Collaboration18},
          Matplotlib \citep{Hunter07},
          NumPy \citep{Harris20}, 
          SciPy \citep{Virtanen20}
          }



%
%
%
%

\bibliographystyle{aasjournal}

\end{document}